\documentclass[fleqn,usenatbib]{mnras}

\usepackage{newtxtext,newtxmath}

\usepackage[T1]{fontenc}
\usepackage{ae,aecompl}

\usepackage{graphicx}	
\usepackage{amsmath}	
\usepackage{amssymb}	
\usepackage{fancyvrb}
\usepackage{subfig}
\usepackage{lscape}
\usepackage{tabularx}

\title[Machine Learning \& Cold Gas Kinematics]{Using machine learning to study the kinematics of cold gas in galaxies}

\author[James M. Dawson et al.]{
James M. Dawson$^{1}$\thanks{E-mail: dawsonj5@cardiff.ac.uk},
Timothy A. Davis$^{1}$, 
Edward L. Gomez$^{1,2}$,
Justus Schock$^{3}$,
\newauthor Nikki Zabel$^{1}$,
and Thomas G. Williams$^{1}$
\\
$^{1}$Cardiff University, School of Physics and Astronomy, The Parade, Cardiff CF24 3AA, UK\\
$^{2}$Las Cumbres Observatory, Suite 102, 6740 Cortona Dr, Goleta, CA 93117, USA\\
$^{3}$RWTH Aachen University, Templergraben 55, 52062 Aachen, Germany
}

\date{Accepted 2019 October 31. Received 2019 October 31; in original form 2019 June 21}

\pubyear{2018}

\begin{document}
\label{firstpage}
\pagerange{\pageref{firstpage}--\pageref{lastpage}}
\maketitle

\begin{abstract}
\noindent Next generation interferometers, such as the Square Kilometre Array, are set to obtain vast quantities of information about the kinematics of cold gas in galaxies. Given the volume of data produced by such facilities astronomers will need fast, reliable, tools to informatively filter and classify incoming data in real time. In this paper, we use machine learning techniques with a hydrodynamical simulation training set to predict the kinematic behaviour of cold gas in galaxies and test these models on both simulated and real interferometric data. Using the power of a convolutional autoencoder we embed kinematic features, unattainable by the human eye or standard tools, into a three-dimensional space and discriminate between disturbed and regularly rotating cold gas structures. Our simple binary classifier predicts the circularity of noiseless, simulated, galaxies with a recall of $85\%$ and performs as expected on observational CO and HI velocity maps, with a heuristic accuracy of $95\%$. The model output exhibits predictable behaviour when varying the level of noise added to the input data and we are able to explain the roles of all dimensions of our mapped space. Our models also allow fast predictions of input galaxies' position angles with a $1\sigma$ uncertainty range of $\pm17^{\circ}$ to $\pm23^{\circ}$ (for galaxies with inclinations of $82.5^{\circ}$ to $32.5^{\circ}$, respectively), which may be useful for initial parameterisation in kinematic modelling samplers. Machine learning models, such as the one outlined in this paper, may be adapted for SKA science usage in the near future. 

\end{abstract}




\section{Introduction}

The age of \textit{Big Data} is now upon us; with the \textbf{S}quare \textbf{K}ilometre \textbf{A}rray (SKA) and \textbf{L}arge \textbf{S}ynoptic \textbf{S}urvey \textbf{T}elescope (LSST) both set to see first light in the mid-2020's.

A key area for \textit{big data} in the next decades will be the studying of the kinematics of cold gas in galaxies beyond our own. This field will rely on interferometers, such as the SKA, thanks to their ability to reveal the morphology and kinematics of the cold gas at high spatial and spectral resolution. Current instruments like the \textbf{A}tacama \textbf{L}arge \textbf{M}illimeter/submillimeter \textbf{A}rray (ALMA) have revolutionised the study of gas in galaxies with their sensitive, high resolution, observations of gas kinematics. However, this field lacks the benefits afforded by fast survey instruments, having long been in an era of \textit{point and shoot} astronomy. As such, large datasets capable of containing global statistics in this research domain have yet to emerge and studies are plagued by slow analytical methods with high user-involvement.

At the time of writing, large-scale radio interferometric surveys such as WALLABY (\citealt{WALLABY}) and APERTIF (\citealt{APERTIF}) are set to begin and will motivate the creation of tools that are scalable to survey requirements. However, these tools will be insufficient for screening objects come the advent of next-generation instruments which are set to receive enormous quantities of data, so large in fact that storing raw data becomes impossible.

In recent times, disc instabilities, feedback, and major/minor mergers have become favoured mechanisms for morphological evolution of galaxies (e.g. \citealt{parry_galaxy_2009,bournaud_black_2011,sales_origin_2012}), the effects of which are visible in their gas kinematics. Therefore, gas kinematics could be used to rapidly identify interesting structures and events suitable for understanding drivers of galaxy evolution (e.g. \citealt{diaz_classifying_2019}). If the kinematics of galaxies can accurately yield information on feedback processes and major/minor merger rates, then astronomers using next generation instruments could develop a better understanding of which mechanisms dominate changes in star formation properties and morphology of galaxies. 
In order to do this we must develop fast, robust, kinematic classifiers. 

Recently, machine learning (ML) has been used successfully in astronomy for a range of tasks including gravitational wave detection (e.g. \citealt{shen_denoising_2017,george_deep_2018,gabbard_matching_2018,zevin_gravity_2017}), exoplanet detection (e.g. \citealt{shallue_identifying_2018}), analysing photometric light curve image sequences (e.g. \citealt{carrasco-davis_deep_2018}), and used extensively in studies of galaxies (e.g. \citealt{dominguez_sanchez_improving_2018,dominguez_sanchez_transfer_2018,dieleman_rotation-invariant_2015,ackermann_using_2018}, \citealt{Bekki2019}).

While using ML requires large data acquisition, training time, resources and the possibility of results that are difficult to interpret, the advantages of using ML techniques over standard tools include (but are not limited to) increased test speed, higher empirical accuracy, and the removal of user-bias. These are all ideal qualities which suit tool-kits for tackling hyper-large datasets. However, the use of ML on longer wavelength millimetre and radio galaxy sources has been absent, with the exception of a few test cases (e.g. \citealt{alger_radio_2018,ma_radio_2018,Andrianomena2019}), with the use of such tests to study the gas kinematics of galaxies being non-existent. It is therefore possible that, in the age of big data, studying gas kinematics with ML could stand as a tool for improving interferometric survey pipelines and encouraging research into this field before the advent of the SKA.  

Cold gas in galaxies that is unperturbed by environmental or internal effects will relax in a few dynamical times. In this state, the gas forms a flat disc, rotating in circular orbits about some centre of potential, to conserve angular momentum. Any disturbance to the gas causes a deviation from this relaxed state and can be observed in the galaxy's kinematics. Ideally therefore, one would like to be able to determine the amount of kinetic energy of the gas invested in circular rotation (the so called \textit{circularity} of the gas; \citealt{sales_origin_2012}). Unfortunately this cannot be done empirically from observations because an exact calculation of circularity requires full six-dimensional information pertaining to the three-dimensional positions and velocities of a galaxy's constituent components. Instead, in the past, astronomers have used approaches such as radial and Fourier fitting routines (e.g. \citealt{Spekkens}, \citealt{Krajnovic}, \citealt{Bloom}) or 2D power spectrum analyses (e.g. \citealt{Grand}) to determine the kinematic regularity of gas velocity fields.

In this work we use a ML model, called a convolutional autoencoder, and a hydrodynamical simulation training set to predict the circularity of the cold interstellar medium in galaxies. We test our resulting model on both simulated test data and real interferometric observations. We use the power of convolutional neural networks to identify features unattainable by the human eye or standard tools and discriminate between levels of kinematic disorder of galaxies. With this in mind, we create a binary classifier to predict whether the cold gas in galaxies exhibit dispersion dominated or disk dominated rotation in order to maximise the recall of rare galaxies with disturbed cold gas.

In \S \ref{section:Autoencoders} we provide the necessary background information for understanding what ML models we use throughout this paper. In \S \ref{section:circularity} we describe the measuring of kinematic regularity of gas in galaxies and how it motivates the use of ML in our work. In \S \ref{Ch.2} we outline our preparation of simulated galaxies into a learnable training set as well as the ML methods used to predict corresponding gas kinematics. In \S \ref{Ch.results} the results of the training process are presented and discussed with a variety observational test cases. Finally, in \S \ref{Ch.conclusion} we explain our conclusions and propose further avenues of research. 

\subsection{Background to convolutional autoencoders}\label{section:Autoencoders}

Convolutional neural networks (CNNs), originally named \textit{neocognitrons} during their infancy (\citealt{fukushima_neocognitron:_1980}), are a special class of neural network (NN) used primarily for classifying multi-channel input matrices, or images. Information is derived from raw pixels, negating the need for a user-involved feature extraction stage; the result being a hyperparametric model with high empirical accuracy. Today, they are used for a range of problems from medical imaging to driverless cars.

A conventional CNN can have any number of layers (and costly operations) including convolutions, max-pooling, activations, fully connected layers, and outputs and often utilise regularisation techniques to reduce overfitting. (For a more in depth background to the internal operations of CNNs we refer the reader to \citealt{krizhevsky_imagenet_2012}). These networks are only trainable (through back propagation) thanks to the use of modern graphics processing units (GPUs; \citealt{steinkraus_using_2005}). It is because of access to technology such as GPUs that we are able to explore the use of ML in a preparatory fashion for instrument science with the SKA in this paper.

A CNN will train on data by minimising the loss between sampled input images and a target variables. Should training require sampling from a very large dataset, training on batches of inputs (also called \textit{mini-batches}) can help speed up training times by averaging the loss between input and target over a larger sample of inputs. Should the network stagnate in minimising the loss, reducing the learning rate can help the network explore a minimum over the parameter space of learnable weights and thus increase the training accuracy. Both of the aforementioned changes to the standard CNN training procedure are used in our models throughout this paper.

An autoencoder is a model composed of two subnets, an \textit{encoder} and a \textit{decoder}. Unlike a standard CNN, during training, an autoencoder learns to reduce the difference between input and output vectors rather than the difference between output vector and target label (whether this be a continuous or categorical set of target classes). In an \textit{undercomplete autoencoder} the encoder subnet extracts features and reduces input images to a constrained number of nodes. This so-called \textit{bottleneck} forces the network to embed useful information about the input images into a nonlinear manifold from which the decoder subnet reconstructs the input images and is scored against the input image using a loss function. With this in mind, the autoencoder works similar to a powerful nonlinear generalisation of principle component analysis (PCA), but rather than attempting to find a lower dimensional hyperplane, the model finds a continuous nonlinear latent surface on which the data best lies. 

Autoencoders have been used, recently, in extra-galactic astronomy for de-blending sources (\citealt{reiman_deblending_2019}) and image generation of active galactic nuclei (AGN; \citealt{ma_radio_2018}).

A convolutional autoencoder (CAE) is very similar to a standard autoencoder but the encoder is replaced with a CNN feature extraction subnet and the decoder is replaced with a transposed convolution subnet. This allows images to be passed to the CAE rather than 1D vectors and can help  interpret extracted features through direct 2D visualisation of the convolution filters. For an intuitive explanation of transposed convolutions we direct the reader to \cite{dumoulin_guide_2016} but for this paper we simply describe a transpose convolution as a reverse, one-to-many, convolution. 

\section{Methodology}\label{Ch.2}

\subsection{Circularity parameter}\label{section:circularity}

As described previously, in order to find and classify kinematic disturbances one would like to measure the circularity of a galaxy's gas disc. For an object composed of point sources (e.g. molecular clouds, stars, etc.), with known positions, masses, and velocities, the circularity measure
\begin{equation}
\kappa = \frac{\text{K}_{\text{rot}}}{\text{K}} \enskip \textrm{where} \enskip \text{K}_{\text{rot}} = \sum^{\text{N}}_{\text{i}=1}\frac{1}{2}\text{m}_{\text{i}} \left(\frac{\text{j}_{\text{z,i}}}{\text{R}_{\text{i}}}\right)^{2} \textrm{and}\enskip \text{K} = \sum^{\text{N}}_{\text{i}=1}\frac{1}{2}\text{m}_{\text{i}}\text{v}_{\text{i}}^{2},
\label{eq:circularity3}
\end{equation}
analyses the fraction of kinetic energy invested in circular, ordered, rotation (\citealt{sales_origin_2012}). Here, K$_\text{rot}$ is a measure of the rotational kinetic energy about some axis and K is the total kinetic energy of the object. m, j, R, and v represent the mass, angular momentum, radius from the centre of rotation, and velocity of each point in an object respectively. Objects with perfectly circular, disk like, rotation have $\kappa = 1$, while objects with either entirely random motion or no motion at all have $\kappa = 0$. 

As $\kappa$ can only be calculated empirically from simulated galaxies, combining ML techniques with simulations will allow us to explore their abilities to learn features that can be used to recover $\kappa$ in observations faster, and more robustly, than by human eye. In fact, $\kappa$ has been used in previous studies to infer the origin of galaxy stellar morphologies (\citealt{sales_origin_2012}) and, more recently, to investigate the kinematics of gas in post starburst galaxies (\citealt{davis_evolution_2019}).

\subsection{EAGLE}\label{section:eagle}

The \textit{Evolution and Assembly of GaLaxies and their Environments} (EAGLE) project\footnote{\label{eagle}\url{http://icc.dur.ac.uk/Eagle/}} is a collection of cosmological hydrodynamical simulations which follow the evolution of galaxies and black holes in a closed volume $\Lambda$ cold dark matter ($\Lambda$CDM) universe. The simulations boast subgrid models which account for physical processes below a known resolution limit (\citealt{schaye_eagle_2015,crain_eagle_2015,the_eagle_team_eagle_2017}). These simulations are able to reproduce high levels of agreement with a range of galaxy properties which take place below their resolution limits (see e.g. \citealt{schaye_eagle_2015}). Each simulation was conducted using smooth particle hydrodynamics, meaning users can directly work with the simulated data in the form of particles, whose properties are stored in output files and a database that can be queried. 

In this paper we make use of these simulations, in conjunction with kinematic modelling tools, to generate a learnable training set. We then probe the use of this training set for transfer learning with the primary goal being to recover kinematic features from generated velocity maps. Using simulations has certain advantages over collecting real data including accessibility, larger sample sizes, and the ability to calculate empirical truths from the data. However, there are drawbacks, including: unproven model assumptions, imperfect physics, and trade-off between resolution and sample size due to computational constraints.

The scripts for reading in data, from the EAGLE project database, were adapted versions of the EAGLE team's pre-written scripts\footnote{\label{eagle_repo}\url{https://github.com/jchelly/read_eagle}}. The original simulations are saved into twenty-nine snapshots for redshifts $z =  0$-$20$ and for this work we utilise snapshot $28$ for \texttt{RefL0025N0376} and \texttt{RefL050N0752} and snapshots 28, 27, 26, and 25 for \texttt{RefL0100N1504} (i.e. redshifts $z = 0$-$0.27$). When selecting galaxies from these snapshots, we set lower limits on the total gas mass ($>1\times10^{9}$ M$_{\odot}$) and stellar mass ($>5\times10^{9}$ M$_{\odot}$) within an aperture size of $30$ kpc around each galaxy's centre of potential (i.e. the position of the most bound particle considering all mass components), in order to exclude dwarf galaxies. In order to select particles which are representative of cold, dense, molecular gas capable of star formation, we only accepted particles with a SFR $>0$ for pre-processing (as described in \S \ref{section:data_prep}). There are many ways to select cold gas in the EAGLE simulations (\citealt{2015MNRAS.452.3815L}) but we use this method for its simplicity as our primary goal is to create a model that is capable of learning low-level kinematic features so as to generalise well in transfer learning tests. The upper radial limit for particle selection of $30$ kpc, from the centre of potential, is in keeping with the scales over which interferometers, such as ALMA, typically observe low-redshift galaxies. It is important that we replicate these scales in order to test our model performance with real data as described in $\S$\ref{section:ALMA_results}. One should note that for future survey instruments, such as the SKA, an alternative scaling via consideration of noise thresholds would be more appropriate. However, as we are particularly interested in the performance of our models with ALMA observations, we instead impose a radial limit for this work. At this stage we also set a lower limit on the number of particles within the $30$ kpc aperture to $>200$. This was to ensure we had enough particles to calculate statistically valid kinematic properties of the galaxies and reduce scaling issues caused by clipping pixels with low brightness when generating velocity maps. With these selection criteria, we work with a set of $14,846$ simulated galaxies.

\subsection{Data preparation}\label{section:data_prep}

Each galaxy was rotated so that their total angular momentum vector was aligned with the positive $z$-axis using the centre of potential (as defined in the EAGLE Database, see \citealt{the_eagle_team_eagle_2017}) as the origin. We then made use of the \texttt{Python} based kinematic simulator \texttt{KinMS}\footnote{\label{kinms}\url{https://github.com/TimothyADavis/KinMSpy}} (KINematic Molecular Simulation) from \cite{davis_black-hole_2013} to turn EAGLE data into mock interferometric observations. \texttt{KinMS} has flexibility in outputting astronomical data cubes (with position, position, and frequency dimensions) and moment maps from various physical parameterisations and has been used for CO molecular gas  modelling in previous work (e.g. \citealt{davis_black-hole_2013}) and for observational predictions from EAGLE (\citealt{davis_evolution_2019}). Using \texttt{KinMS} we generate simulated interferometric observations of galaxies directly from their 3D particle distributions.

Thanks to the controllable nature of the EAGLE data, we have the ability to generate millions of images from just a handful of simulations by using combinations of rotations and displacements of thousands of simulated galaxies per snapshot. This flexibility also has the added benefit of naturally introducing data augmentation for boosting the generalising power of an ML algorithm. For any given distance projection, galaxies were given 8 random integer rotations in position-angle ($0^{\circ}\leq \theta_{\text{pos}}<360 ^{\circ}$) and inclination ($5^{\circ}\leq\phi_{\text{inc}}\leq85^{\circ}$). Each galaxy is displaced such that they fill a $64^{\prime\prime}\times64^{\prime\prime}$ mock velocity map image in order to closely reflect the field of view (FOV) when observing CO(1-0) line emission with ALMA. We define the displacement of each simulated galaxy in terms of their physical size and desired angular extent. Each galaxy's radius is given as the $98$th percentile particle distance from its center of potential in kpc. We use this measurement, rather than the true maximum particle radius, to reduce the chance of selecting sparsely populated particles for calculating displacement distances, as they can artificially scale down galaxies.

The EAGLE galaxies were passed to \texttt{KinMS} to create cubes of stacked velocity maps, with fixed mock beam sizes of $\text{bmaj} = 3^{\prime\prime}$, ready for labelling. Each cube measured $64\times64\times8$ where $64\times64$ corresponds to the image dimensions (in pixels) and $8$ corresponds to snapshots during position-angle and inclination rotations. The median physical scale covered by each pixel across all image cubes in a representative sample of our training set is $0.87$ kpc. It should be noted that we set all non-numerical values or infinities to a constant value, as passing such values to an ML algorithm will break its training. We adopt $0$ $\text{km\,s}^{-1}$ as our constant (similarly to \citealt{diaz_classifying_2019}) to minimise the the background influencing feature extraction. Our training set has a range in blank fraction (i.e. the fraction of pixels in images with blank values set to 0 $\text{km\,s}^{-1}$) of 0.14 to 0.98, with a median blank fraction of 0.52. Figure \ref{Figure:EAGLE_examples} shows simulated ALMA observations of galaxies when using \texttt{KinMS} in conjunction with particle data from the EAGLE simulation \texttt{RefL0025N0376}. 

\subsection{Simulating noise}\label{section:noise}

Often it is useful to observe the performance of ML models when adding noise to the input data, in order to test their robustness and their behavioural predictability. In one of our tests, we seeded the mock-EAGLE-interferometric-datacubes with Gaussian distributed noise of mean $\mu = 0$ and standard deviation 
\begin{equation} 
\sigma = \frac{1}{\text{S/N}}\left(\frac{1}{N}\sum\limits_{\text{c}=0}^{\text{c}=\text{N}} \text{I}_{\text{max,c}}\right), 
\end{equation} 
i.e. some fraction, $\frac{1}{\text{S/N}}$, of the mean maximum intensity, I$_{\text{max}}$, of each cube-channel, $c$, containing line emission. The resulting noisy data cubes are then masked using \textit{smooth masking}, a method that is representative of how one would treat a real data cube (\citealt{Dame2011}).
An intensity weighted moment one map is then generated in \texttt{KinMS} from the masked cube as 
\begin{equation}
\text{M}_{1} = \frac{\int(\text{v})\text{I}_{\text{v}} \text{dv}}{\int \text{I}_{\text{v}} \text{dv}} = \frac{\sum(\text{v})\text{I}_{\text{v}}}{\sum\text{I}_{\text{v}}},
\label{eq:moment1}
\end{equation}
where I$_{\text{v}}$ is the observed intensity in a channel with known velocity $\text{v}$, before being normalised into the range of $-1$ to $1$.  

Noise presents a problem when normalising images into the preferred range. Rescaling, using velocities beyond the range of real values in a velocity map (i.e. scaling based on noise), will artificially scale down the true values and thus galaxies will appear to exhibit velocities characteristic of lower inclinations. We clip all noisy moment 1 maps at a fixed $96$th percentile level, before normalising, in order to combat this effect. Note that this choice of clipping at the $96$th percentile level is arbitrarily based on a handful of test cases and represents no specific parameter optimisation. Although simple, this likely reflects the conditions of a next generation survey in which clipping \textit{on the fly} will be done using a predetermined method globally rather than optimising on a case by case basis.

\subsection{Labelling the training set}

\begin{figure}
\includegraphics[width=\linewidth]{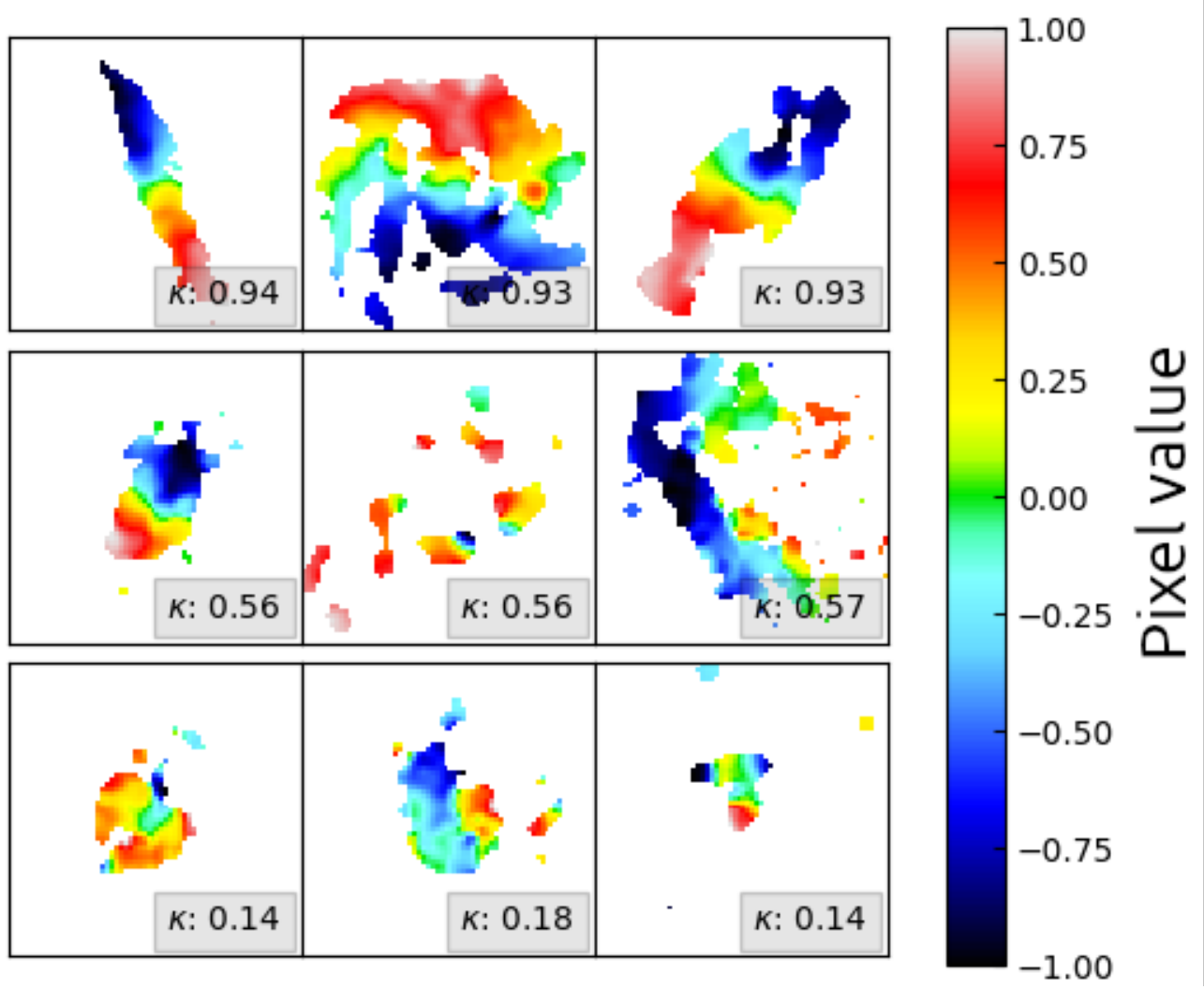}
\caption{Random exemplar velocity maps for the noiseless EAGLE dataset. Rows of increasing order, starting from the bottom of the figure, show galaxies of increasing $\kappa$. The $\kappa$ for each galaxy is shown in the bottom right of the frame in a grey box. Each galaxy has randomly selected position angle and inclination and the colourbar indicates the line of sight velocities, which have been normalised into the range $-1$ to $1$ and subsequently denoted as \textit{pixel values}. The images have dimensions of $64\times64$ pixels in keeping with the size of input images to our models in this paper, as described in \S\ref{section:training}. One can easily see the changes in velocity field from $\kappa\sim1$ to $\kappa\sim0$ as galaxies appear less disk-like with more random velocities.}
\label{Figure:EAGLE_examples}
\end{figure}

Each galaxy, and therefore every cube, is assigned a label in the continuous range of $0$ to $1$ corresponding to the level of ordered rotation, $\kappa$, of that galaxy.

In Figure \ref{Figure:EAGLE_examples}, the difference between levels of $\kappa$ is clear in both structure and velocity characteristics, with low $\kappa$ galaxies exhibiting less regular structures and more disturbed velocity fields than high $\kappa$ galaxies. 

\begin{figure}
\includegraphics[width=\linewidth]{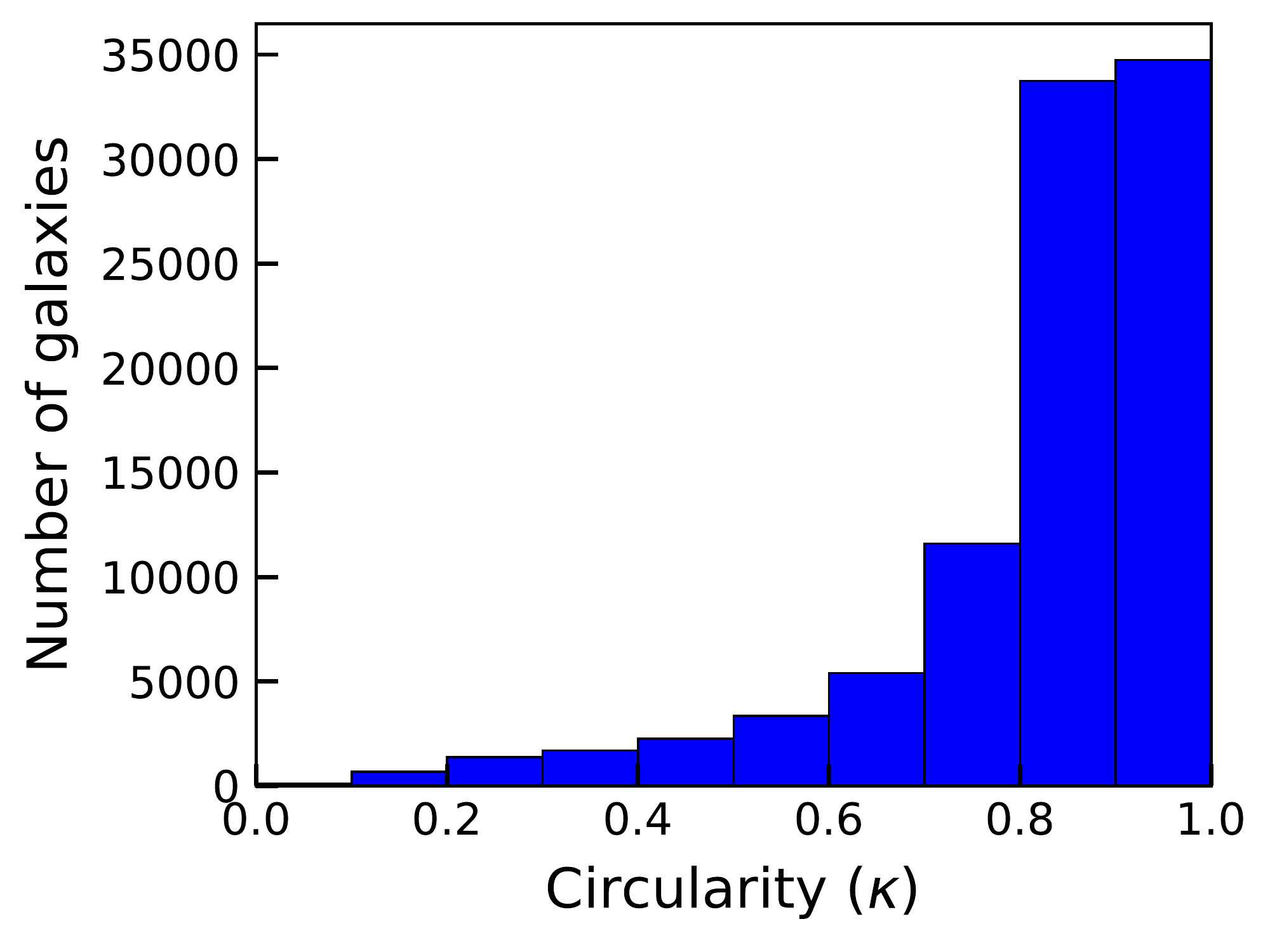}
\caption{A histogram of $\kappa$ labelled galaxies in the noiseless EAGLE training set. Galaxies have been binned in steps of $\delta\kappa = 0.1$ for visualisation purposes but remain continuous throughout training and testing. The distribution of $\kappa$ is heavily imbalanced, showing that more galaxies exhibit a $\kappa$ closer to 1 than 0.}
\label{Figure:kappas}
\end{figure}

Figure \ref{Figure:kappas} shows the distribution of $\kappa$ in our training set. It is clear that our training set is heavily imbalanced with a bias towards the presence of high $\kappa$ galaxies. Additionally, as $\kappa$ approaches one, the possible variation in velocity fields decreases as there are limited ways in which one can create orderly rotating disk-like structures. However, our dataset contains a surplus of galaxies as $\kappa$ approaches one. Therefore, if one were to randomly sample from our dataset, for training an ML model, then the model would undoubtedly overfit to high $\kappa$ images. This is a common problem in ML particularly with outlier detection models whose objectives are to highlight the existence of rare occurrences. In $\S$\ref{section:training} we describe our solution for this problem with the use of weighted sampling throughout training to balance the number of galaxies with underrepresented $\kappa$ values seen at each training epoch.
 
\begin{figure*}
\centering
\subfloat[Encoder subnet]{%
\includegraphics[width=\linewidth]{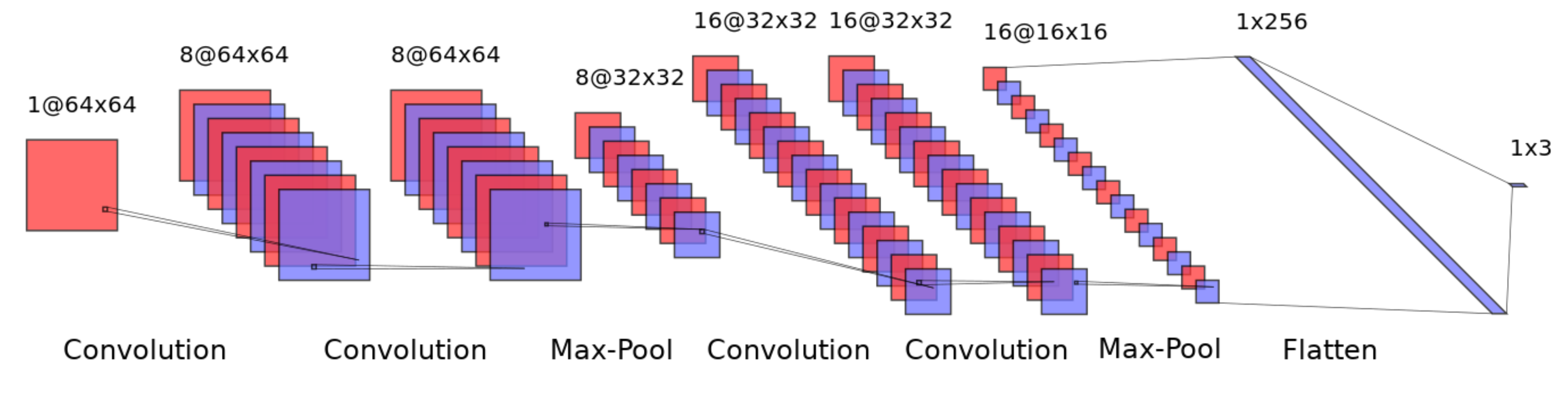}
\label{Figure:encoder}%
}\qquad
\subfloat[Decoder subnet]{%
\includegraphics[width=\linewidth]{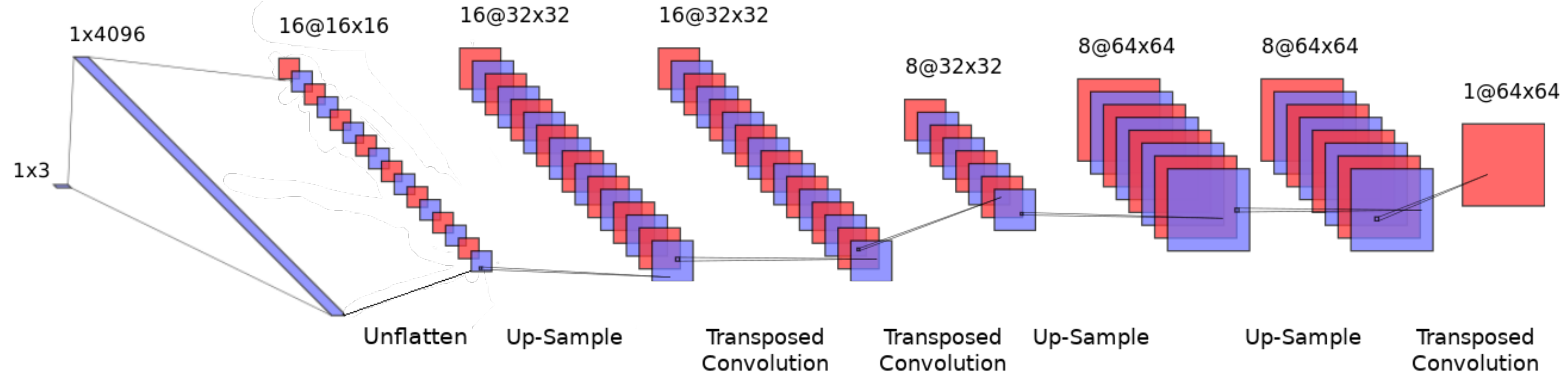}%
\label{Figure:decoder}%
}
\caption{Illustration of the CAE architecture used in this paper. The encoder subnet (top) makes use of a series of convolutions and max-pooling operations to embed input image information into 3 latent dimensions. The decoder subnet (bottom) recovers the input image using transposed convolutions and up-sampling layers. The output of the encoder is passed to the decoder during training but throughout testing only the encoder is used map velocity maps into latent space.}
\label{Figure:autoencoder}
\end{figure*}

\subsection{Model training: Rotationally invariant case}\label{section:training}

\begin{figure}
\includegraphics[width=\linewidth]{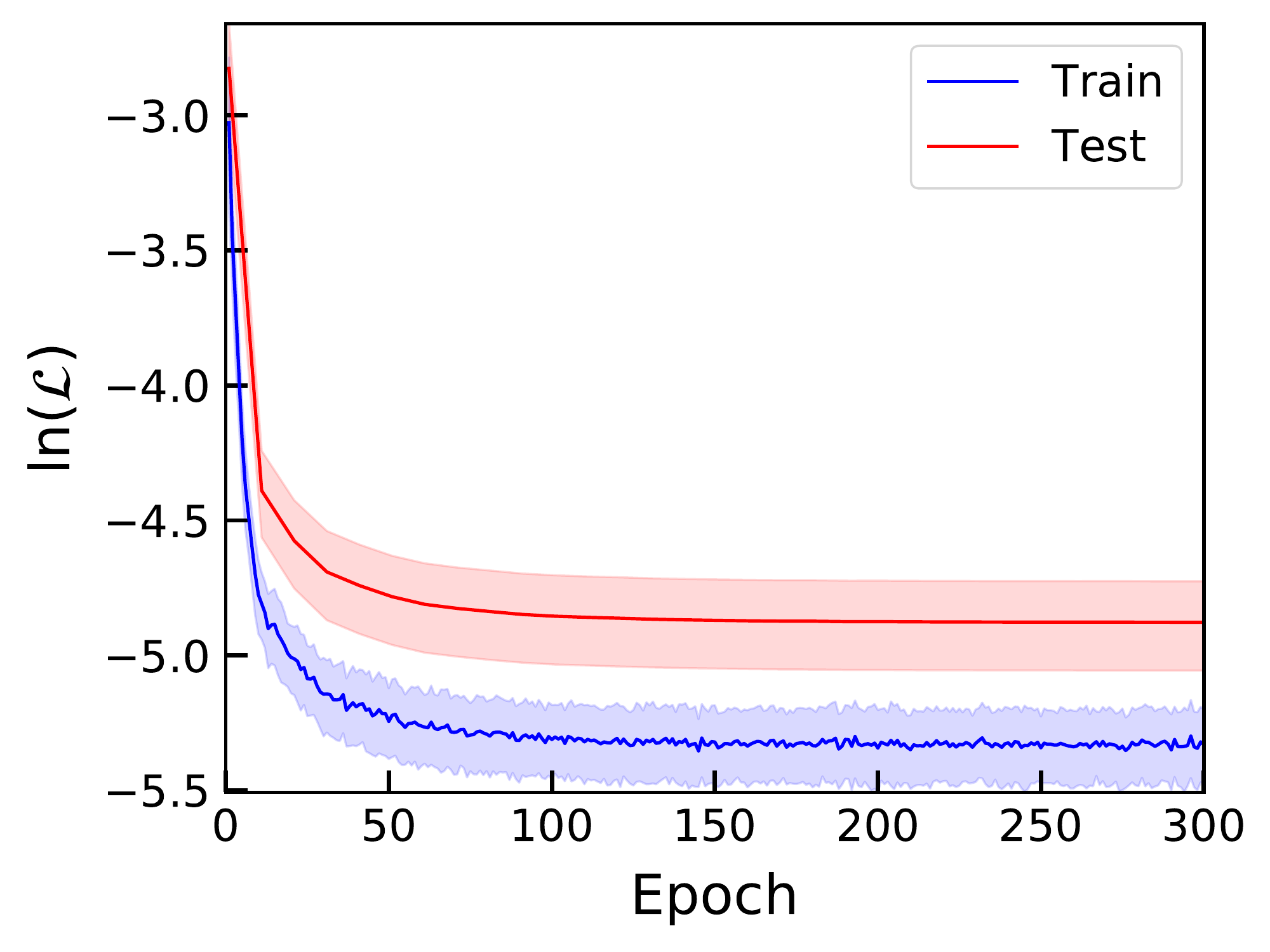}
\caption{Training the CAE on noiseless EAGLE velocity maps. Solid lines show the natural log mean MSE loss and solid colour regions show $1\sigma$ spread at any given epoch. In order to reduce computational time, the test accuracy is evaluated every $10$th epoch. We see smooth convergence of our CAE throughout training with no turn over of the test accuracy indicating that our model did not overfit to the training data.}
\label{Figure:training}
\end{figure}

In this section we describe the creation and training of a convolutional autoencoder to embed $\kappa$ into latent space and build a binary classifier to separate galaxies with $\kappa$ above and below $0.5$. Note that $0.5$ is an arbitrarily chosen threshold for our classification boundary but is motivated by the notion of separating ordered from disturbed gas structures in galaxies.

In order to construct our ML model, we make use of \texttt{PyTorch}\footnote{\label{pytorch}\url{http://pytorch.org/}} 0.4.1, an open source ML library capable of GPU accelerated tensor computation and automatic differentiation (\citealt{paszke_automatic_2017}). Being grounded in \texttt{Python}, \texttt{PyTorch} is designed to be linear and intuitive for researchers with a C99 API backend for competitive computation speeds. We use \texttt{PyTorch} due to its flexible and user friendly nature for native \texttt{Python} users.

A visual illustration of the CAE architecture is shown in Figure \ref{Figure:autoencoder} and described in Table \ref{table:AE} in more detail. The model follows no hard structural rules and is an adaption of standard CNN models. The decoder structure is simply a reflection of the encoder for simplicity. This means our CAE is unlikely to have the most optimised architecture and we propose this as a possible avenue for improving on the work presented in this paper. The code developed for this paper is available on GitHub\footnote{\label{GitHub1}\url{https://github.com/SpaceMeerkat/CAE/releases/tag/v1.0.0}} as well as an ongoing development version\footnote{\label{GitHub2}\url{https://github.com/SpaceMeerkat/CAE}}. 

The CAE is trained for 300 epochs (with a batch size of 32) where one epoch comprises a throughput of 6400 images sampled from the training set. We do this to reduce the memory load throughout training given such a large training set. Images are selected for each mini-batch using a weighted sampler which aims to balance the number of images in each $\kappa$ bin of width $\delta\kappa=0.1$. Inputs are sampled with replacement allowing multiple sampling of objects to prevent under-filled bins. The model uses a mean squared error (MSE) loss, 
\begin{equation}
\mathcal{L} = \frac{1}{\text{N}} \sum\limits_{\text{i}=0}^{\text{N}} \left(f\left(\text{x}_{\text{i}}\right)-\text{y}_{\text{i}}\right)^{2}, 
\end{equation}
for evaluating the error between input and output images and weights are updated through gradient descent. N, $f(\text{x})$, and y denote the batch size, model output for an input x, and target respectively. We use an adaptive Adam learning rate optimiser (\citealt{kingma_adam:_2014}), starting with a learning rate of $0.001$ which halves every $30$ epochs; this helps to reduce stagnation in the model accuracy from oversized weight updates. In Figure \ref{Figure:training} we see that the model has converged well before the $300^{\text{th}}$ epoch and observe no turn-over of the test MSE loss, which would indicate overfitting.

\begin{figure*}
\includegraphics[width=\linewidth]{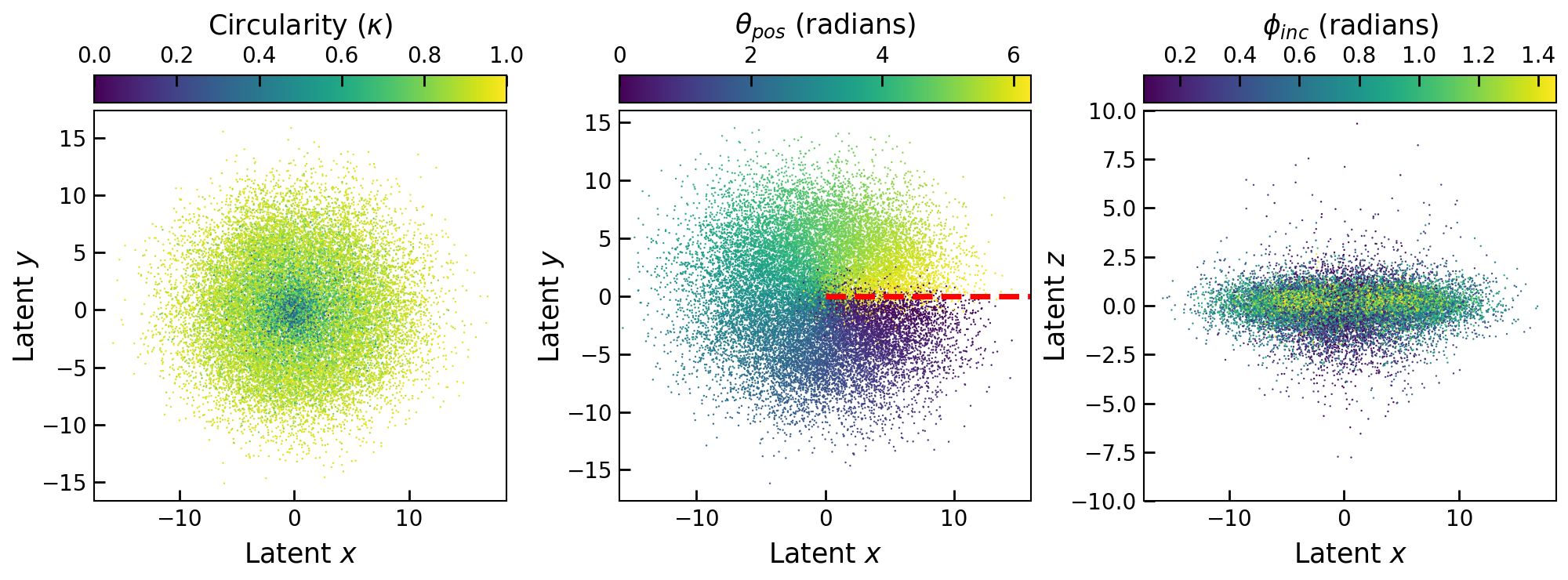}
\caption{Noiseless eagle test data in 3D latent space. All subplots show the same latent structure but coloured differently by: true $\kappa$ (left), true position angle ($\theta_{\text{pos}}$, middle), and true inclination ($\phi_{\text{inc}}$, right).  It is clear from the right subplot that low $\kappa$ galaxies lie close to the $z=0$ region. $\theta_{\text{pos}}$ is very neatly encoded in the clockwise angle around the latent $z$-axis. The red dashed line indicates the positive latent $x$ axis from which $\theta_{pos}$ is measured. $\phi_{\text{inc}}$ appears to be encoded in a much more complex fashion than $\kappa$ and $\theta_{\text{pos}}$.}
\label{Figure:latent_3d}
\end{figure*}

The CAE learned to encode input images to 3 dimensional latent vectors. Further testing showed that any higher compression, to lower dimensions, resulted in poor performance for the analyses described in $\S$\ref{Ch.results} and compression to higher dimensions impaired our ability to directly observe correlations between features and latent positions with no improvement to the model's performance. We use \texttt{scikit-learn}'s\footnote{\label{scikit-learn}\url{https://scikit-learn.org/}} principal component analysis (PCA) function on these vectors to  rotate the latent space so that it aligns with one dominant latent axis, in this case the $z$ axis. As seen in Figure \ref{Figure:latent_3d}, the 3 dimensional latent space contains structural symmetries which are not needed when attempting to recover $\kappa$ (but are still astrophysically useful; see \S\ref{sec:pos_ang}). Because of this, the data is folded around the z and x axes consecutively to leave a 2-dimensional latent space devoid of structural symmetries with dimensions |z| and $\sqrt{x^{2}+y^{2}}$ from from which we could build our classifier (see \S\ref{section:ALMA_results}).

\begin{figure}
\includegraphics[width=\linewidth]{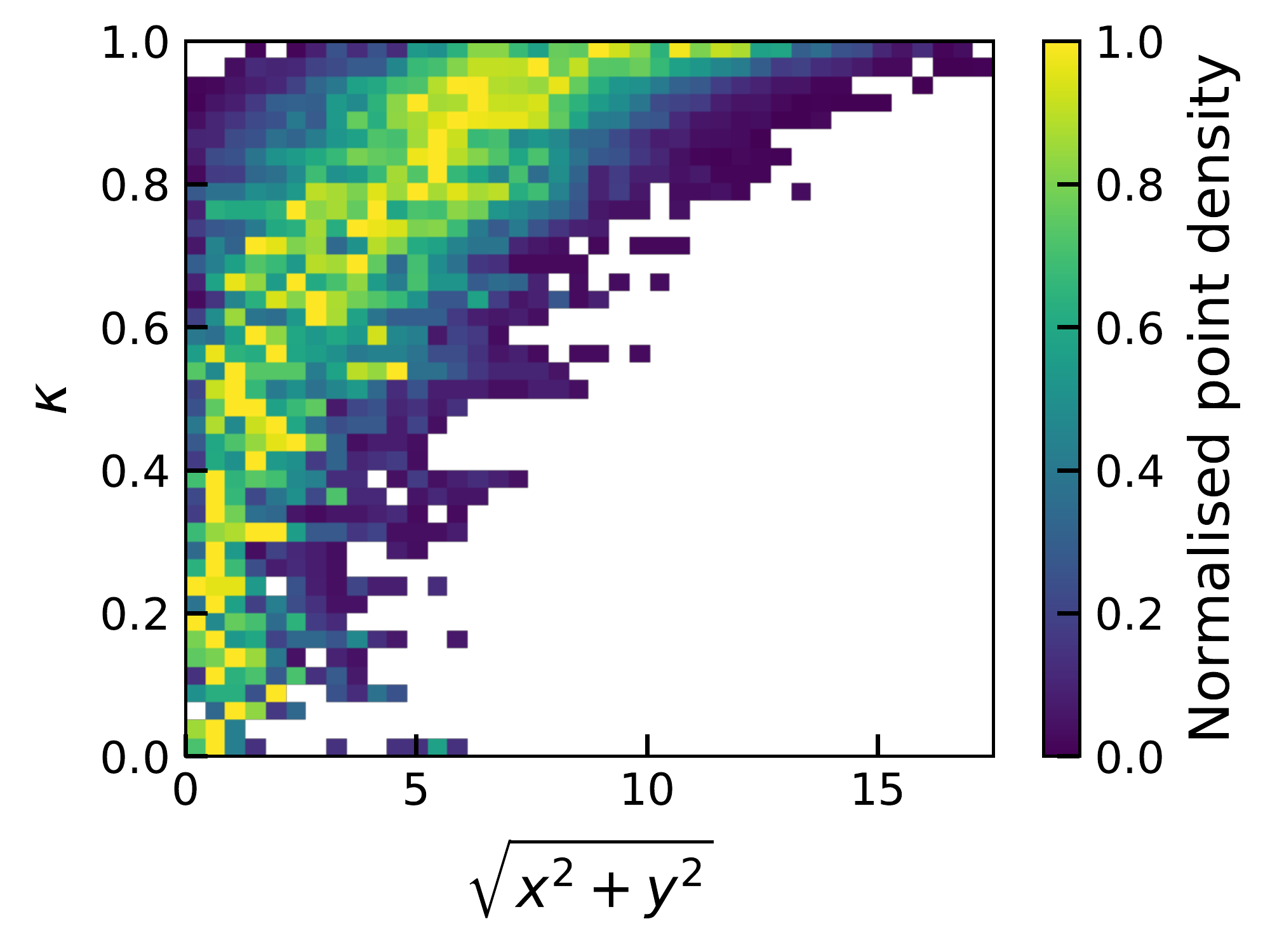}
\caption{2D histogram of $\kappa$ against latent position for noiseless EAGLE test data. Pixels are coloured by point density normalised such that the point density in each row lies in the range 0 to 1. We see a very clear relationship between $\kappa$ and latent position but also a high spread of latent positions occupied by high $\kappa$ galaxies, making a regression task to recover $\kappa$ from our encoding difficult.}
\label{Figure:point_density}
\end{figure}

\begin{figure}
\includegraphics[width=\linewidth]{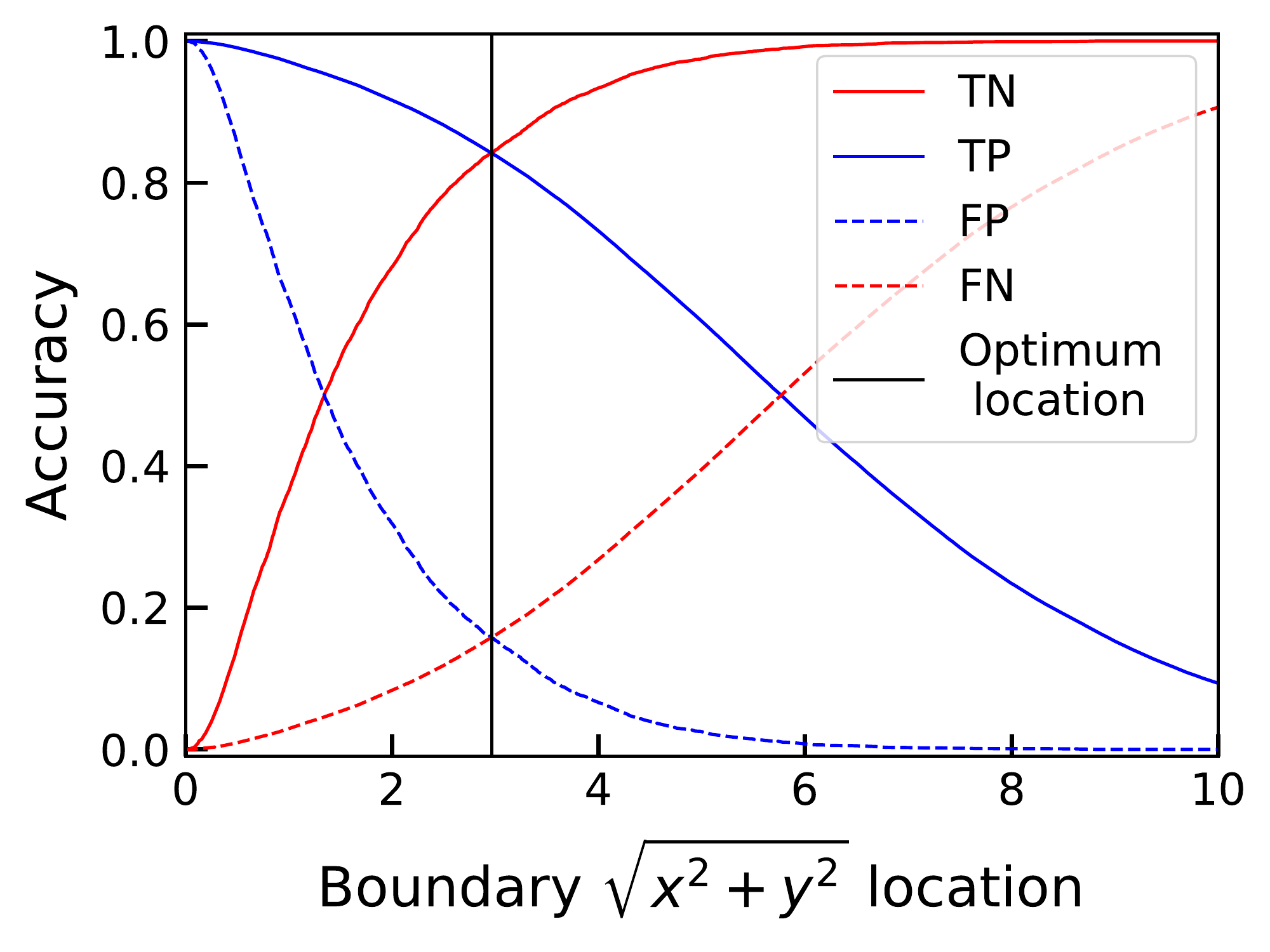}
\caption{The observed change in all four components of a confusion matrix when changing the boundary line $x$-location. The optimal position for a binary classification is chosen as the intersection of TP and TN lines, which is identical to the location at the intersection of FP and FN lines. We observe smooth changes to the TN, TP, FP, and FN lines as the boundary line location changes, showing that both target populations are well clustered.}
\label{Figure:TPTN}
\end{figure}

Having tested multiple classifiers on the 2D latent space (such as high order polynomial and regional boundary approaches), we find that a simple vertical boundary line is best at separating the galaxies whose $\kappa$ are greater than or less than $0.5$. This is highlighted in Figure \ref{Figure:point_density}, where we see the spread on latent positions taken up by different $\kappa$ galaxies makes a regression to recover $\kappa$ too difficult. In order to optimise the boundary line location, we measure the true positive (TP), true negative (TN), false positive (FP) and false negative (FN) scores when progressively increasing the boundary line's $x$ location. The intersection of TP and TN lines (and therefore the FP and FN lines) in Figure \ref{Figure:TPTN} indicates the optimal position for our boundary, which is at $\sqrt{x^{2}+y^{2}}=2.961\pm0.002$. The smoothness of the lines in Figure \ref{Figure:TPTN} show how the two $\kappa$ populations are well structured. If the two populations were clumpy and overlapping, one would observe unstable lines as the ratio of positive and negatively labelled galaxies constantly shifts in an unpredictable manner. 

\section{Results and discussion}\label{Ch.results}

\begin{table}
\caption{Proportions of high and low $\kappa$ labelled images in both training and test sets for the noiseless EAGLE dataset.}
\centering
\begin{tabular}{lccc}
\cline{2-4}
\multicolumn{1}{r}{} & \multicolumn{3}{c}{Number of images}  \\ \cline{2-4} 
Dataset              & $\kappa>0.5$   & $\kappa<0.5$ & Total \\ \cline{1-4}
Training             & 88840 ($94\%$) & 6144 ($6\%$) & 94984 \\ \cline{1-4}
Test                 & 22224 ($93\%$) & 1560 ($7\%$) & 23784 \\ \cline{1-4}
\end{tabular}
\label{table:datasets}
\end{table}

\begin{figure*}
\subfloat[Noiseless EAGLE train]{%
  \includegraphics[width=0.324\linewidth]{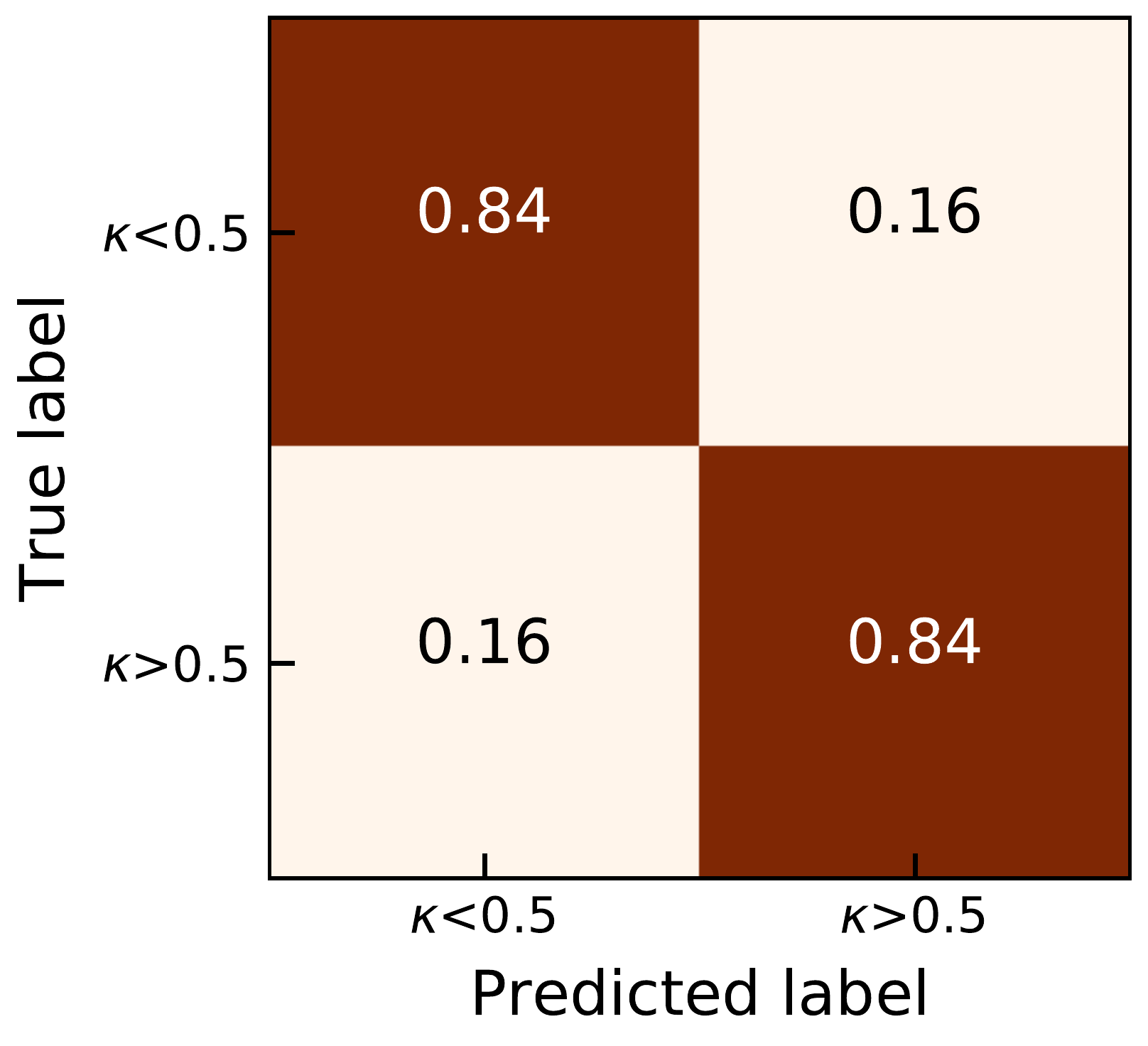}
  \label{Figure:CM1}%
}\qquad
\subfloat[Noiseless EAGLE test]{%
  \includegraphics[width=0.3\linewidth]{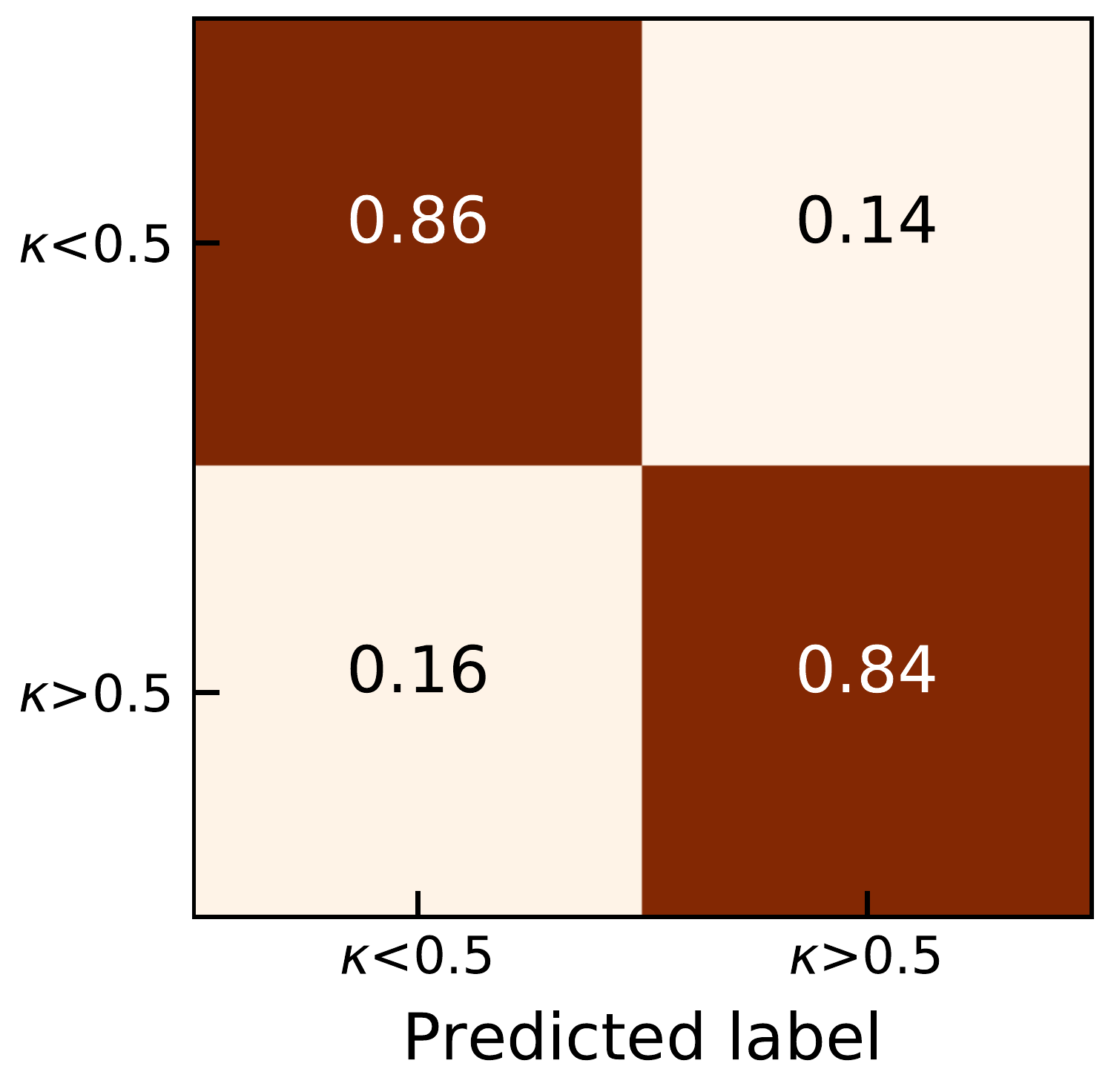}%
  \label{Figure:CM2}%
}\qquad
\subfloat[Noisy EAGLE test]{%
  \includegraphics[width=0.3\linewidth]{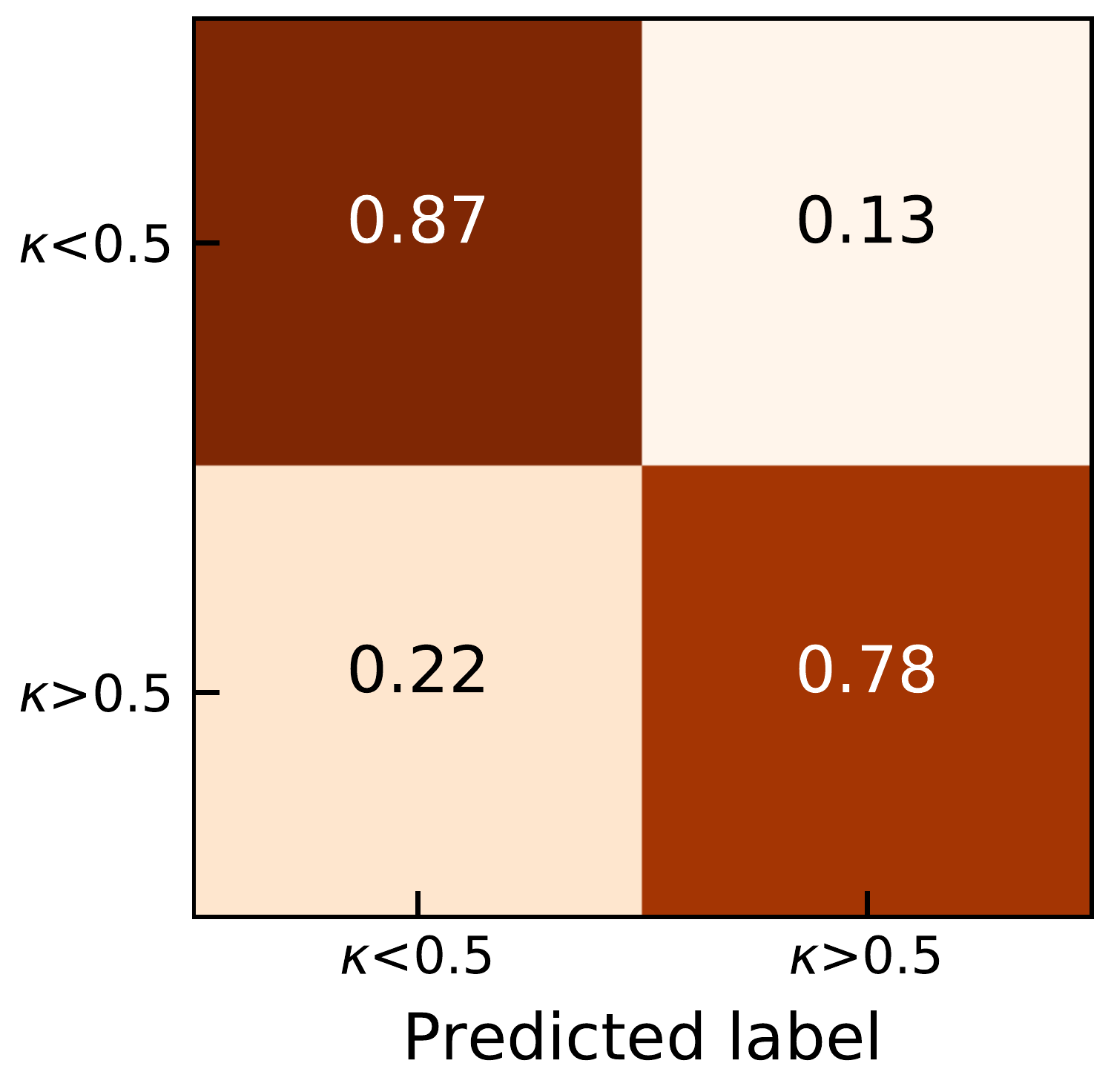}%
  \label{Figure:CM3}%
}
\caption{Normalised confusion matrix showing the performance of the classifier when testing the \ref{Figure:CM1} noiseless EAGLE training set, \ref{Figure:CM2} noiseless EAGLE test set (seeded with Gaussian noise with $\frac{1}{\text{S/N}} = \frac{1}{10}$ and masking at 3 times the RMS level of line free regions), and \ref{Figure:CM3} noisy EAGLE test set. The mean recall scores are $84\%$, $85\%$, and $82.5\%$ respectively.}
\label{Figure:CM_results}
\end{figure*}

\subsection{Test case I: Noiseless EAGLE data}

The number of high and low $\kappa$ labelled images, in both the training and test sets, for the noiseless EAGLE dataset are shown in Table \ref{table:datasets}. Figure \ref{Figure:CM1} shows the classification accuracy on the noiseless EAGLE training set. The TP and TN accuracy scores are unsurprisingly identical given the method used to find the optimal boundary in $\S$\ref{section:training} was designed to achieve this (see intersection points in Figure \ref{Figure:TPTN}). The classifier has a mean training recall of $84\%$ for both classes.

Figure \ref{Figure:CM2} shows the confusion matrix when testing the noiseless EAGLE test set using our boundary classifier. We see that the model performs slightly better than when tested on the training set, suggesting that the model did not overfit to the training data and is still able to encode information on $\kappa$ for unseen images.

\subsection{Test case II: Noisy EAGLE data}\label{sec:adding_noise}

Figure \ref{Figure:CM3} shows the results of classifying noisy EAGLE test data with $\text{S/N} = 10$ and masking at $3$ times the RMS level (see $\S$\ref{section:noise} for details). Note that this is a simple test case and places no major significance on the particular level of S/N used. The introduction of noise has a clear and logical, yet arguably minor, impact on the classifier's accuracy.
The combination of adding noise followed by using an arbitrary clipping level causes test objects to gravitate towards the low $\kappa$ region in latent space. This should come as no surprise as $\kappa$ correlates with ordered motion; therefore, any left over noise from the clipping procedure, which itself appears as disorderly motions and structures in velocity maps, anti-correlates with $\kappa$ causing a systematic shift towards the low $\kappa$ region in latent space. 

One could reduce this shifting to low $\kappa$ regions in several ways. (1) Removing low S/N galaxies from the classification sample. (2) For our test cases we used a single absolute percentile level for smooth clipping noise; using levels optimised for cases on a one-by-one basis will prevent over-clipping. (3) If one were to directly sample the noise properties from a specific instrument, seeding the simulated training data with this noise before retraining an CAE would cause a systematic shift in the boundary line, mitigating a loss in accuracy. It should also be noted that we have not tested the lower limit of S/N for which it is appropriate to use our classifier but instead we focus on demonstrating the effects of applying noise clipping globally across our test set under the influence of modest noise.

\subsection{Test case III: ALMA data}\label{section:ALMA_results}

We tested 30 velocity maps of galaxies observed with ALMA to evaluate the performance of the classifier on real observations. Given that we used \texttt{KinMS} to tailor the simulated velocity maps to closely resemble observations with ALMA we expect similar behaviour as seen when testing the simulated data. For our test sample we use an aggregated selection of 15 velocity maps from the mm-\textbf{W}ave \textbf{I}nterferometric \textbf{S}urvey of \textbf{D}ark \textbf{O}bject \textbf{M}asses (WISDOM) CO(1-0, 2-1, and 3-2) and 15 CO(1-0) velocity maps from the \textbf{A}LMA \textbf{Fo}rnax \textbf{C}luster \textbf{S}urvey (AlFoCS, \citealt{zabel_alma_2019}). We classify each galaxy, by eye, as either disturbed or regularly rotating (see Table \ref{table:ALMA}) in order to heuristically evaluate the classifier's performance.

\begin{landscape}
\begin{figure}
\vspace{1.5cm}
\centering
\includegraphics[width=\linewidth,keepaspectratio]{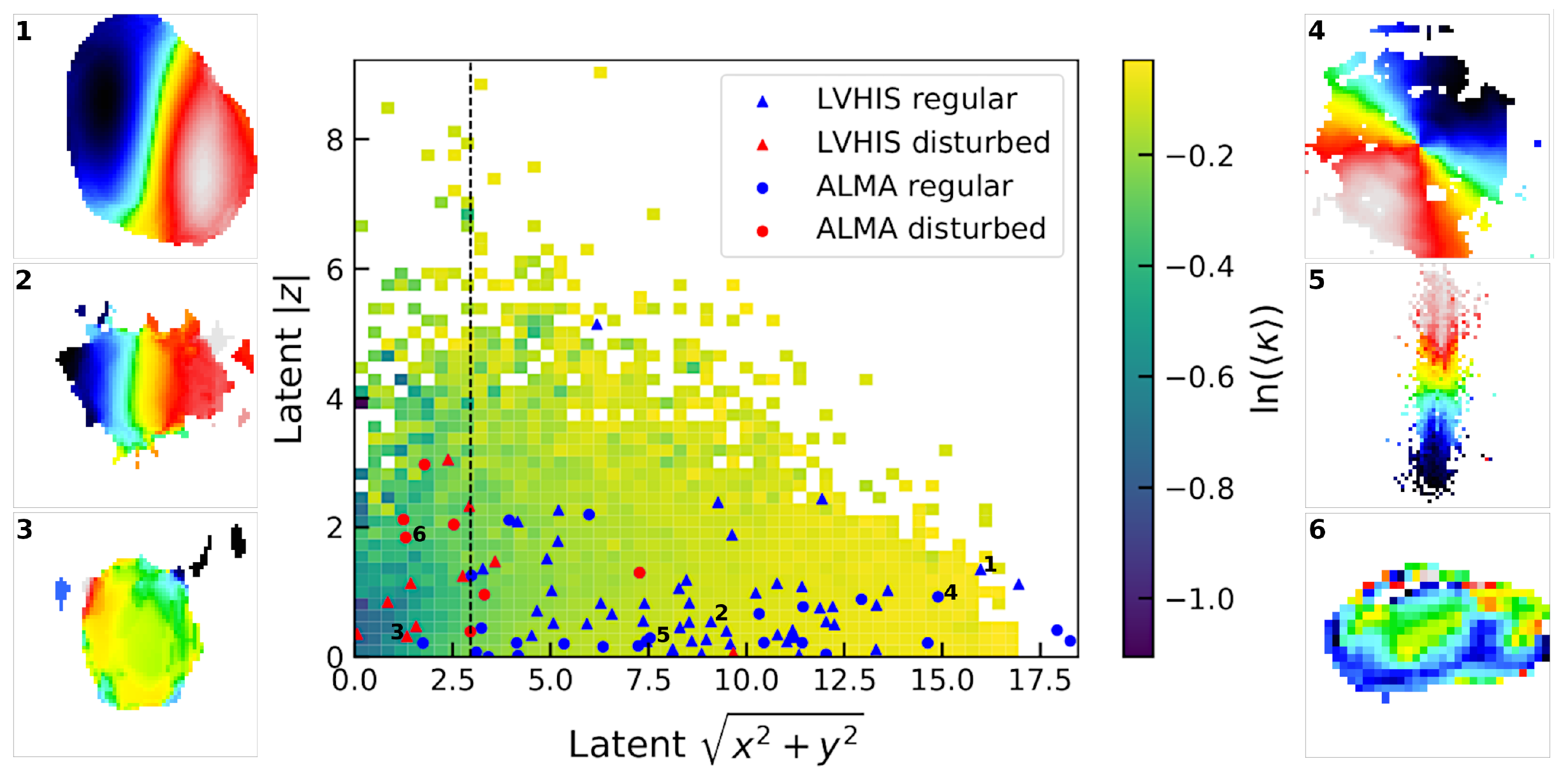}
\caption{Folded latent space positions of noiseless EAGLE galaxies (coloured 2D histogram), ALMA galaxies (circular markers), and LVHIS galaxies (triangular markers); the suspected low $\kappa$ galaxies are coloured red and suspected high $\kappa$ galaxies coloured blue. The EAGLE points are coloured by average value of $\kappa$ in each 2D bin and the grey dashed line shows the classifier boundary between $\kappa<0.5$ and $\kappa>0.5$ objects. 8 of the 10 suspected low $\kappa$ LVHIS galaxies are classified as $\kappa<0.5$ and 5 of the 7 suspected low $\kappa$ ALMA galaxies are classified as $\kappa<0.5$. In both cases we see only 1 misclassification far from the boundary line and low $\kappa$ region. A selection of 3 LVHIS and 3 ALMA galaxies are shown to the left and right of the central plot respectively, their locations in latent space indicated by black numbers. Images are scaled to $64\times64$ pixels with values normalised between $-1$ to $1$, because of this images are shown as a close representation of CAE inputs before latent encoding. For illustration purposes the backgrounds have been set to \texttt{nan} whereas the CAE would instead see these regions as having a value of $0$.}
\label{Figure:latent_ALMA_LVHIS}
\end{figure}
\end{landscape}

Figure \ref{Figure:latent_ALMA_LVHIS} shows the positions of all ALMA galaxies (round markers) in our folded latent space, once passed through the CAE. Of the $30$ galaxies, $6$ ($20\%$) are classified as $\kappa<0.5$; this higher fraction, when compared to the fraction of low $\kappa$ galaxies in the simulated test set, is likely due to the high number of dwarf galaxies, with irregular H$_2$ gas, targeted in AlFoCS.

 We find one false positive classification close to the classification boundary and one false positive classification far from the classification boundary. The false negative classification of NGC1351A can be explained by its disconnected structure and edge-on orientation (see \citealt{zabel_alma_2019}; Figure B1). Since low $\kappa$ objects appear disconnected and widely distributed among their velocity map fields of view, it is understandable why NGC1351A has been misclassified as a disturbed object. It should be noted that upon inspection the false positive classification of FCC282 can be attributed to the appearance of marginal rotation in the galaxy. We see evidence of patchy high $\kappa$ galaxies residing closer to the classification boundary than non-patchy examples. This may indicate a relationship between patchiness and latent positioning. The classifier performs with an accuracy of $90\%$ when compared to the predictions by human eye. Of the $30$ galaxies observed with ALMA, $6$ ($20\%$) are classified as low $\kappa$ and of the $23$ ($77\%$) galaxies identified by eye as likely to be high $\kappa$ galaxies, only one was misclassified as low $\kappa$.

\subsection{Test case IV: LVHIS data}\label{sec:HI_results}

In order to test the robust nature of the classifier, we used it to classify velocity maps of HI velocity fields from the the \textbf{L}ocal \textbf{V}olume \textbf{HI} \textbf{S}urvey (LVHIS; \citealt{koribalski_local_2018}). This is an important test as it determines the applicability of the classifier to HI line emission observations, the same emission that the SKA will observe. As described in $\S$\ref{section:data_prep}, the EAGLE training set was designed to reflect observations with ALMA, making this transfer learning test a good opportunity to evaluate the model's ability to generalise to unseen data containing different systematic characteristics.

Rather than moment masking the data cubes, like in $\S$\ref{sec:adding_noise}, each cube is clipped at some fraction of the RMS (calculated in line free channels) to mimic the noise removal processes used in generating velocity maps in the LVHIS database. All galaxies whose positions could not be found using the \texttt{Python} package \texttt{astroquery}\footnote{\label{astroquery}\url{https://astroquery.readthedocs.io/en/latest/}} (searching the SIMBAD Astronomical Database\footnote{\label{simbad}\url{http://simbad.u-strasbg.fr/simbad/}}), or whose HI structures were clearly misaligned with the true galaxy centres, were omitted from further testing. This was to prevent misclassification based on pointing error which correlates with features of disorderly rotation to the CAE and would artificially increase the FN rate. This left $61$ galaxies (see Table \ref{table:LVHIS}) from which velocity maps were made and passed through the CAE. Finally, where images were not $64 \times 64$ pixels, we used PyTorch's \texttt{torch.nn.functional.interpolation} function (in bilinear mode) to rescale them up or down to the required dimensions prior to clipping.

The latent positions of all HI galaxies are shown in Figure \ref{Figure:latent_ALMA_LVHIS} (triangular markers). Of the $61$ galaxies, $8$ ($13\%$) are classified as low $\kappa$. By eye, we identified $10$ galaxies in the LVHIS which are likely to be definitively classified as $\kappa < 0.5$ (see Table \ref{table:LVHIS}). Of these $10$ candidates, $8$ were correctly identified as $\kappa < 0.5$, $1$ is observed as very close to the to the classification boundary and $1$ is unquestionably misclassified. 

\subsection{Recovering position angle}\label{sec:pos_ang}

\begin{figure}
\includegraphics[width=\linewidth]{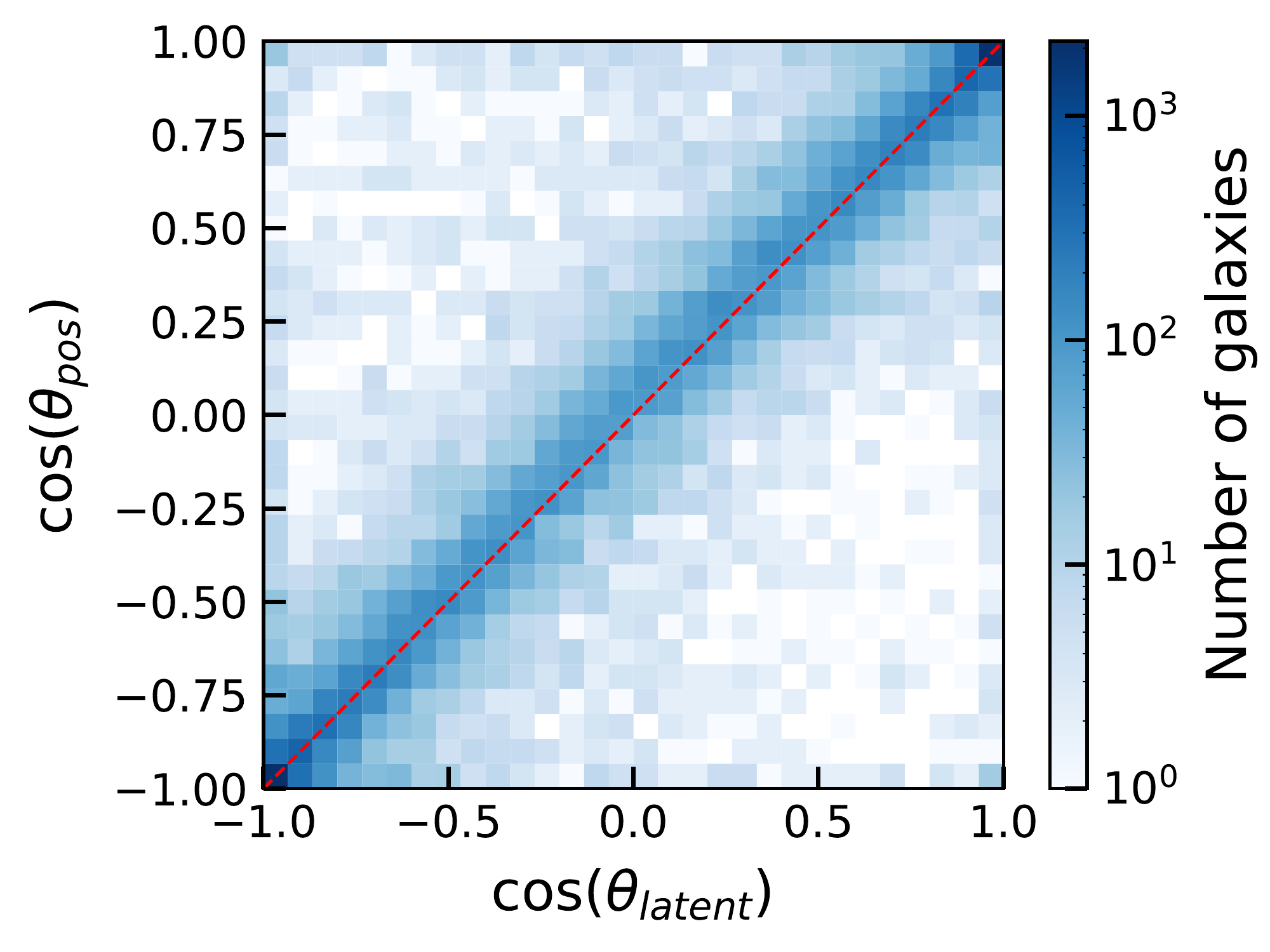}
\caption{2D histogram showing the predicted position angles for the noiseless EAGLE test set against their true position angles. The red dashed line shows the 1:1 line along which all data would lie for a perfect predictor of position angle.}
\label{Figure:position_angle}
\end{figure}

Scientists who wish to model the kinematics of galaxies often require initial estimates for parameters such as position angle, inclination, mass components, radial velocity profiles etc. Given that position angle is clearly encoded in our latent $xy$ plane (see Figure \ref{Figure:latent_3d}), it is possible to return predicted position angles with associated errors. This could prove useful for fast initial estimates of $\theta_{\text{pos}}$ for scientists requiring them for kinematic modelling. We define the predicted position angle, $\theta_{\text{latent}}$, as the clockwise angle between the positive latent $x$-axis and the position of data points in the latent $xy$ plane. We removed the systematic angular offset, $\delta\theta$, between the positive latent $x$-axis and the true position angle ($\theta_{\text{pos}}$) $= 0^{\circ}$ line by rotating the latent positions by the median offset, found to be $\delta\theta\sim36.6^{\circ}$, and subtracting an additional $180^{\circ}$. In the now rotated frame, $\theta_{\text{latent}}$ is defined as $\tan^{-1}\left(\frac{y}{x}\right)$, where $x$ and $y$ are the latent $x$ and $y$ positions of each galaxy (see Figure \ref{Figure:position_angle}). We calculated errors on the resulting predictions of $\theta_{\text{latent}}$ by taking the standard deviation of residuals between $\theta_{\text{latent}}$ and $\theta_{\text{pos}}$. 

\begin{figure}
\includegraphics[width=\linewidth]{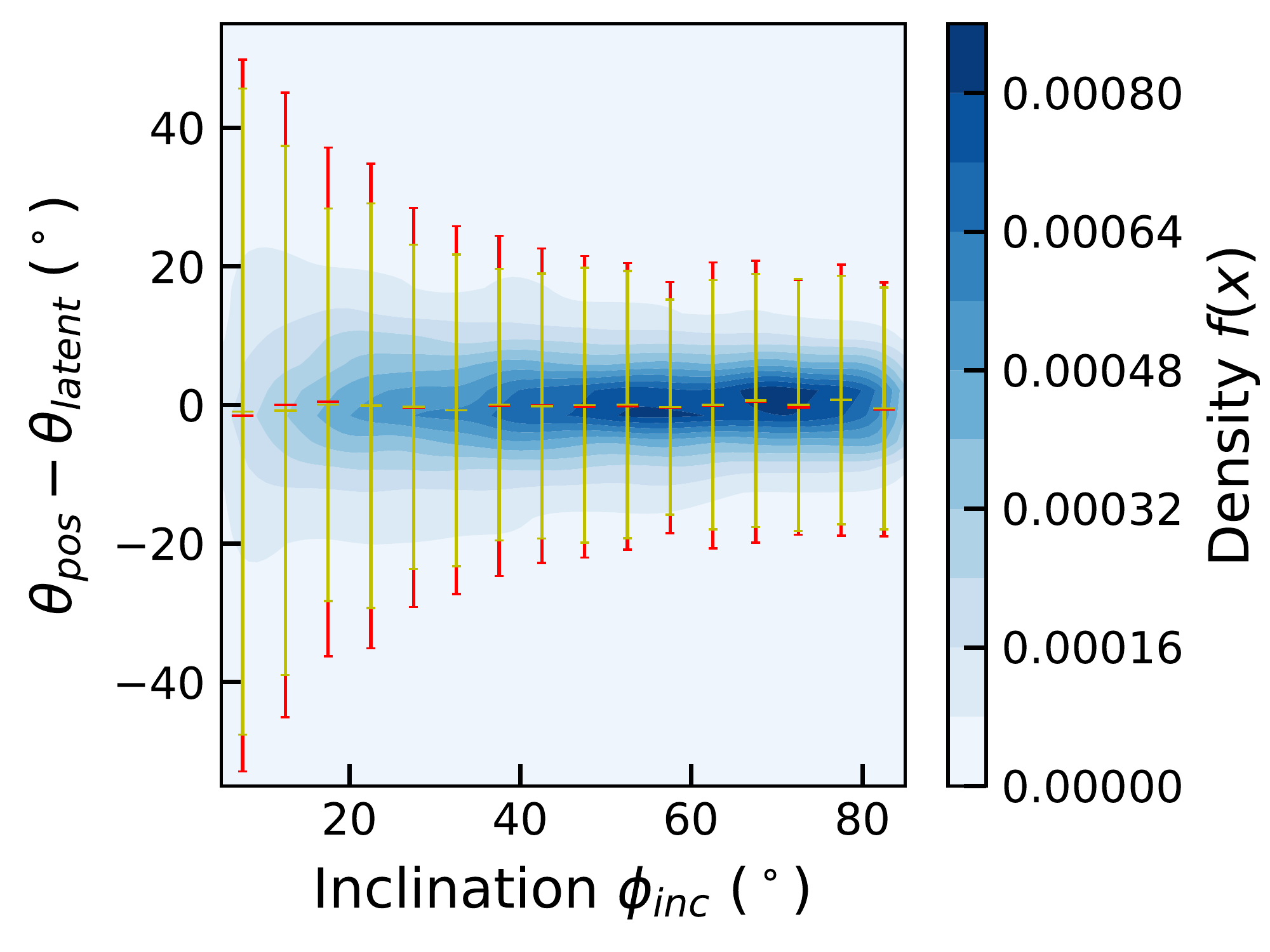}
\caption{Kernel density estimation of error in $\theta_{\text{pos}}$ against inclination for noiseless EAGLE test data (yellow error bars) and noisy EAGLE test data (red error bars). Coloured contours show the 2D probability density, central horizontal line markers show the mean error in $\theta_{\text{pos}}$ in bins of width $\delta\theta_{\text{pos}} = 5^{\circ}$. The error bars show the standard error in each bin.}
\label{Figure:pos_vs_inc}
\end{figure}

We repeated this procedure for the noisy EAGLE data, with $\text{S/N}=10$, the results of which are also shown in Figure \ref{Figure:pos_vs_inc} with red error bars. We can see that the recovery of $\theta_{\text{pos}}$ is still well constrained at higher inclinations with only a slight increase in the error most notably at lower inclinations (see Figure \ref{Figure:pos_vs_inc}. We see that at higher inclinations the error in predicted $\theta_{\text{pos}}$ is better constrained than for lower inclinations. This should come as no surprise as the ellipticity of galaxies and the characteristic shape of their isovels are gradually lost as a galaxy approaches lower inclinations thus making it more difficult to calculate $\theta_{\text{pos}}$. During further testing we also observe reduced errors on position angles when limiting to higher $\kappa$ test galaxies.

It should be noted that our method for recovering $\theta_{\text{pos}}$ is not the only one. Other kinematic fitting routines exist for this purpose including \texttt{fit\_kinematic\_pa} (\citealt{Krajnovi_kinemetry}) and the radon transform method (\citealt{Stark_radon}). These methods likely have higher accuracy than seen here, as our network was not optimised for the recovery of $\theta_{pos}$. Bench marking an ML model against existing ones, as a dedicated standalone mechanism for recovering $\theta_{\text{pos}}$, is an avenue for future research.

Given that there is such a strong overlap in $z$-positions occupied by different galaxy inclinations, we were unable to recover the inclinations of galaxies in the simple manner as for $\theta_{\text{pos}}$. However, from visualising the distribution of inclinations against latent-$z$ position, we are confident that inclination plays a part in latent positioning of galaxies. Because of this we are confident in our understanding of all 3 latent dimensions that the CAE has learned.



\section{Conclusions}\label{Ch.conclusion}

We have shown that it is possible to use ML to encode high dimensional ($64\times64$ pixels) velocity maps into 3 dimensions, while retaining information on galaxy kinematics, using convolutional autoencoders. We have successfully recovered the level of ordered rotation of galaxies using a simple binary classifier, from a multitude of test sets including simulated EAGLE velocity maps, ALMA velocity maps, and HI survey velocity maps. When testing real observational data, we see a clustering of low $\kappa$ galaxies towards the origin and around the classification boundary, in line with our understanding of our folded two dimensional latent space. Our tests on simulated data show a mean recall of $85\%$ when attempting to recover the circularity parameter as well as $90\%$ and $97\%$ heuristic accuracy when recovering the circularity parameter for galaxies observed with ALMA and as part of LVHIS respectively. We have managed to mitigate the problems associated with a heavily imbalanced training set by using both weighted sampling during training and balancing the true positive and true negative accuracy scores when constructing our classifier. In addition to recovering information on the ordered rotation characteristics of galaxies, we have also been successful in providing estimates on position angle from the full 3D latent positions of velocity maps with associated errors. These will be useful for initial guesses at $\theta_{\text{pos}}$ for kinematic modelling routines in related work.  

We were able to show our classifier's positive performance when testing LVHIS data. This outcome is important for two reasons: (1) it shows the robustness of the classifier when making the transition from simulated to real data of different origins and (2) it shows that using machine learning to study the kinematics of HI sources is likely possible and therefore applicable to SKA science. 

Recovering inclinations, $\phi_{\text{inc}}$, of galaxies was not possible using our CAE due to the high overlap in latent $z$ positions for the entire range of $\phi_{\text{inc}}$. However, the spread of $z$ positions occupied by galaxies at mid-range inclinations was considerably less than at lower inclinations, indicating that while $\phi_{\text{inc}}$ is not recoverable, we are confident that it is partly responsible for the positions of galaxies in the latent $z$-axis. Therefore, we have a rational understanding of what information all three latent dimensions are encoding from the input images. This makes our model predictable and logical in how it behaves when seeing input data. This understanding is often missing in CNN style networks, and especially in deep learning models. 

The main caveat with this work pertains to the use of percentages in our maximum likelihood function when calculating the optimal boundary line for the binary classifier. This makes our classifier independent of the underlying distribution of high and low $\kappa$ galaxies in an attempt to maximise the recall of both classes. The means our classifier will work well in situations where both classes are more equally distributed (such as galaxy clusters). However, one should take care when testing heavily imbalanced datasets where, although the dataset has been drastically thinned of high $\kappa$ galaxies, it is likely that the user will still need to examine the low $\kappa$ classification set for contaminants.

As demonstrated by \cite{diaz_classifying_2019}, using a combination of morphology and kinematics for classification purposes improves performance over using only one attribute. Therefore, a logical improvement on our work would be using a branched network or an ensemble of networks which use both moment zero and moment one maps to make predictions on kinematic properties. Our models rely on using maps of galaxies which are centred on their centres of potential (i.e. the position of the most bound particle); therefore, our classifier is sensitive to the choice of centre of potential proxy. This is undeniably an issue for on-the-fly surveys where the centre of potential of a target is estimated rather than empirically calculable. Therefore, including information such as intensity maps may allow re-centring based on observed characteristics rather than archived pointings for improving the classifiers performance. We see this as the most lucrative avenue for improving our models in the future.

Performing operations on a velocity map, as we have done in this work, means we are working several levels of abstraction away from the raw datacubes that future instruments, such as the SKA, will create. Therefore improvements could be made on our methods to analyse the effects of encoding datacubes into latent space rather than velocity maps. CNNs have long been capable of performing operations on multi-channel images, making this avenue of research possible and useful in reducing the need for heavy processing of raw datacubes before processing with ML algorithms as we have done in the work. 

\section*{Acknowledgements}

The authors would like to thank the anonymous reviewer for their useful comments which improved this manuscript.

This paper has received funding by the Science and Technology Facilities Council (STFC) as part of the Cardiff, Swansea \& Bristol Centre for Doctoral Training.

We gratefully acknowledge the support of NVIDIA Corporation with the donation of a Titan Xp GPU used for this research. 

This research made use of Astropy\footnote{http://www.astropy.org}, a community-developed Python package for Astronomy \citep{astropy_collaboration_astropy:_2013,astropy_collaboration_astropy_2018}.

This paper makes use of the following ALMA data: \\
ADS/JAO.ALMA\#2013.1.00493.S ADS/JAO.ALMA\#2015.1.00086.S ADS/JAO.ALMA\#2015.1.00419.S ADS/JAO.ALMA\#2015.1.00466.S ADS/JAO.ALMA\#2015.1.00598.S ADS/JAO.ALMA\#2016.1.00437.S ADS/JAO.ALMA\#2016.1.00839.S ADS/JAO.ALMA\#2015.1.01135.S ADS/JAO.ALMA\#2016.1.01553.S ADS/JAO.ALMA\#2016.2.00046.S

ALMA is a partnership of the ESO (representing its member states), NSF (USA), and NINS (Japan), together with the NRC (Canada), NSC, ASIAA (Taiwan), and KASI (Republic of Korea), in cooperation with the Republic of Chile. The Joint ALMA Observatory is operated by the ESO, AUI/NRAO, and NAOJ.

JD wishes to thank Dr. Freeke van de Voort, from the Max Planck Institute for Astrophysics, for her help in querying the EAGLE database and understanding the circularity parameter used throughout this paper.

We thank the PyTorch community for their assistance in learning the ropes of implementing and training neural networks with PyTorch. PyTorch is an optimized, open source, tensor library for deep learning using GPUs and CPUs.






\bibliographystyle{mnras}
\bibliography{Paper_1_citations.bib}


\appendix
\section{Information on test galaxies}

\begin{table*}
\begin{tabular}{cllcccc}
\hline
\hline
        & \multicolumn{1}{c}{Layer} & \multicolumn{1}{c}{Layer Type}              & Units/number of filters & Size    & Padding & Stride \\ \hline
Encoder & Input                     & Input                    &       & (64,64) &         &        \\ \cline{2-7} 
        & Conv1                     & Convolutional            & 8     & (3,3)   & 1       & 1      \\ \cline{2-7} 
        & ReLU                      & Activation               &       &         &         &        \\ \cline{2-7} 
        & Conv2                     & Convolutional            & 8     & (3,3)   & 1       & 1      \\ \cline{2-7}
        & ReLU                      & Activation               &       &         &         &        \\ \cline{2-7} 
        & MaxPool                   & Max-pooling              &       & (2,2)   &         & 1      \\ \cline{2-7} 
        & Conv3                     & Convolutional            & 16    & (3,3)   & 1       & 1      \\ \cline{2-7} 
        & ReLU                      & Activation               &       &         &         &        \\ \cline{2-7} 
        & Conv4                     & Convolutional            & 16    & (3,3)   & 1       & 1      \\ \cline{2-7} 
        & ReLU                      & Activation               &       &         &         &        \\ \cline{2-7} 
        & MaxPool                   & Max-pooling              &       & (2,2)   &         & 1      \\ \cline{2-7} 
        & Linear                    & Fully-connected          & 3    &         &         &        \\ \hline
Decoder & Linear                    & Fully-connected          & 3    &         &         &        \\ \cline{2-7} 
        & Up                        & Partial inverse max-pool &       & (2,2)   &         & 1      \\ \cline{2-7} 
        & ReLU                      & Activation               &       &         &         &        \\ \cline{2-7} 
        & Trans1                    & Transposed Convolution   & 16    & (3,3)   & 1       & 1      \\ \cline{2-7} 
        & ReLU                      & Activation               &       &         &         &        \\ \cline{2-7} 
        & Trans2                    & Transposed Convolution   & 16    & (3,3)   & 1       & 1      \\ \cline{2-7} 
        & Up                        & Partial inverse max-pool &       & (2,2)   &         & 1      \\ \cline{2-7} 
        & ReLU                      & Activation               &       &         &         &        \\ \cline{2-7} 
        & Trans3                    & Transposed Convolution   & 8     & (3,3)   & 1       & 1      \\ \cline{2-7} 
        & ReLU                      & Activation               &       &         &         &        \\ \cline{2-7} 
        & Trans4                    & Transposed Convolution   & 8     & (3,3)   & 1       & 1      \\ \cline{2-7} 
        & Ouput                     & Output                   &       & (64,64) &         &        \\ \hline
\end{tabular}
\caption{Architecture for our autoencoder, featuring both encoder and decoder subnets. The decoder is a direct reflection of the encoder's structure.}
\label{table:AE}
\end{table*}

\begin{table*}
\begin{tabular}{rcccc}
\hline
\hline
\multicolumn{1}{c}{OBJECT ID} & 
\multicolumn{1}{c}{SURVEY} & 
\multicolumn{1}{c}{\begin{tabular}[c]{@{}c@{}}Author Prediction \\ (disturbed=0, regular=1)\end{tabular}} & 
\multicolumn{1}{c}{\begin{tabular}[c]{@{}c@{}}Model Prediction \\ ($\kappa<0.5$=0, $\kappa>0.5$=1) \end{tabular}}& 
\multicolumn{1}{c}{\begin{tabular}{@{}c@{}c@{}}Heuristic Result \\ (TP=true positive, FP=false positive \\ 
TN=true negative, FN=false negative) \end{tabular}} \\ \hline
ESO358-G063   & AlFoCS & 1 & 1 & TP \\
ESO359-G002   & AlFoCS & 0 & 0 & TN\\
FCC207        & AlFoCS & 1 & 1 & TP\\
FCC261        & AlFoCS & 0 & 0 & TN\\
FCC282        & AlFoCS & 0 & 1 & FP\\
FCC332        & AlFoCS & 0 & 0 & TN\\
MCG-06-08-024 & AlFoCS & 0 & 0 & TN\\
NGC1351A      & AlFoCS & 1 & 0 & TN\\
NGC1365       & AlFoCS & 1 & 1 & TP\\
NGC1380       & AlFoCS & 1 & 1 & TP\\
NGC1386       & AlFoCS & 1 & 1 & TP\\
NGC1387       & AlFoCS & 1 & 1 & TP\\
NGC1436       & AlFoCS & 1 & 1 & TP\\
NGC1437B      & AlFoCS & 1 & 1 & TP\\
PGC013571     & AlFoCS & 0 & 1 & FP\\
NGC0383       & WISDOM & 1 & 1 & TP\\
NGC0404       & WISDOM & 0 & 0 & TN\\
NGC0449       & WISDOM & 1 & 1 & TP\\
NGC0524       & WISDOM & 1 & 1 & TP\\
NGC0612       & WISDOM & 1 & 1 & TP\\
NGC1194       & WISDOM & 1 & 1 & TP\\
NGC1574       & WISDOM & 1 & 1 & TP\\
NGC3368       & WISDOM & 1 & 1 & TP\\
NGC3393       & WISDOM & 1 & 1 & TP\\
NGC4429       & WISDOM & 1 & 1 & TP\\
NGC4501       & WISDOM & 1 & 1 & TP\\
NGC4697       & WISDOM & 1 & 1 & TP\\
NGC4826       & WISDOM & 1 & 1 & TP\\
NGC5064       & WISDOM & 1 & 1 & TP\\
NGC7052       & WISDOM & 1 & 1 & TP\\ \hline
\end{tabular}
\caption{ALMA galaxies selected from the WISDOM and AlFoCS surveys. WISDOM targets have beam major axes ranging from $2.4^{\prime\prime}$ to $6.7^{\prime\prime}$ with a mean of $4.4^{\prime\prime}$ and pixels/beam values ranging from $2.42$ to $6.68$ with a median value of $4.46$. ALL WISDOM targets have channel widths of $2$ $\text{km\,s}^{-1}$ bar one target which has a channel width of $3$ $\text{km\,s}^{-1}$. AlFoCS targets have beam major axes ranging from $2.4^{\prime\prime}$ to $3.3^{\prime\prime}$ with a mean of $2.9^{\prime\prime}$ and pixels/beam values ranging from $5.25$ to $7.85$ with a median value of $6.46$. AlFoCS targets have channel widths ranging from $9.5$ to $940$ $\text{km\,s}^{-1}$, with a median channel width of $50$ $\text{km\,s}^{-1}$. Of all 30 galaxies in the test set, 7 were identified by eye as most likely to be classified as $\kappa < 0.5$ and their associated model predictions are shown. $27$ ($90\%$) of the galaxies are classified as predicted by human eye. NGC1351A is the only false negative classification owing to its disconnected structure and edge on orientation.}
\label{table:ALMA}
\end{table*}

\begin{table*}
\begin{tabular}{rrccc}
\hline
\hline
\multicolumn{1}{c}{LVHIS ID} & 
\multicolumn{1}{c}{OBJECT ID} & 
\multicolumn{1}{c}{\begin{tabular}[c]{@{}c@{}}Author Prediction \\ (disturbed=0, regular=1)\end{tabular}} & 
\multicolumn{1}{c}{\begin{tabular}[c]{@{}c@{}}Model Prediction \\ ($\kappa<0.5$=0, $\kappa>0.5$=1)\end{tabular}} & 
\multicolumn{1}{c}{\begin{tabular}{@{}c@{}c@{}}Heuristic Result \\ (TP=true positive, FP=false positive \\ 
TN=true negative, FN=false negative) \end{tabular}}\\ \hline
LVHIS 001 & ESO 349-G031    & 1 & 1 & TP\\
LVHIS 003 & ESO 410-G005    & 0 & 1 & FP\\
LVHIS 004 & NGC 55          & 1 & 1 & TP\\
LVHIS 005 & NGC 300         & 1 & 1 & TP\\
LVHIS 007 & NGC 247         & 1 & 1 & TP\\
LVHIS 008 & NGC 625         & 1 & 1 & TP\\
LVHIS 009 & ESO 245-G005    & 1 & 1 & TP\\
LVHIS 010 & ESO 245-G007    & 0 & 0 & TN\\
LVHIS 011 & ESO 115-G021    & 1 & 1 & TP\\
LVHIS 012 & ESO 154-G023    & 1 & 1 & TP\\
LVHIS 013 & ESO 199-G007    & 1 & 1 & TP\\
LVHIS 015 & NGC 1311        & 1 & 1 & TP\\
LVHIS 017 & IC 1959         & 1 & 1 & TP\\
LVHIS 018 & NGC 1705        & 1 & 1 & TP\\
LVHIS 019 & ESO 252-IG001   & 1 & 1 & TP\\
LVHIS 020 & ESO 364-G?029   & 1 & 1 & TP\\
LVHIS 021 & AM 0605-341     & 1 & 1 & TP\\
LVHIS 022 & NGC 2188        & 1 & 1 & TP\\
LVHIS 023 & ESO 121-G020    & 1 & 1 & TP\\
LVHIS 024 & ESO 308-G022    & 1 & 1 & TP\\
LVHIS 025 & AM 0704-582     & 1 & 1 & TP\\
LVHIS 026 & ESO 059-G001    & 1 & 1 & TP\\
LVHIS 027 & NGC 2915        & 1 & 1 & TP\\
LVHIS 028 & ESO 376-G016    & 1 & 1 & TP\\
LVHIS 029 & ESO 318-G013    & 1 & 1 & TP\\
LVHIS 030 & ESO 215-G?009   & 1 & 1 & TP\\
LVHIS 031 & NGC 3621        & 1 & 1 & TP\\
LVHIS 034 & ESO 320-G014    & 1 & 1 & TP\\
LVHIS 035 & ESO 379-G007    & 1 & 1 & TP\\
LVHIS 036 & ESO 379-G024    & 0 & 0 & TN\\
LVHIS 037 & ESO 321-G014    & 1 & 1 & TP\\
LVHIS 039 & ESO 381-G018    & 1 & 1 & TP\\
LVHIS 043 & NGC 4945        & 1 & 1 & TP\\
LVHIS 044 & ESO 269-G058    & 1 & 1 & TP\\
LVHIS 046 & NGC 5102        & 1 & 1 & TP\\
LVHIS 047 & AM 1321-304     & 0 & 0 & TN\\
LVHIS 049 & IC 4247         & 0 & 1 & FP\\
LVHIS 050 & ESO 324-G024    & 1 & 1 & TP\\
LVHIS 051 & ESO 270-G017    & 1 & 1 & TP\\
LVHIS 053 & NGC 5236        & 1 & 1 & TP\\
LVHIS 055 & NGC 5237        & 1 & 1 & TP\\
LVHIS 056 & ESO 444-G084    & 1 & 1 & TP\\
LVHIS 057 & NGC 5253        & 0 & 0 & TP\\
LVHIS 058 & IC 4316         & 0 & 0 & TP\\
LVHIS 060 & ESO 325-G?011   & 1 & 1 & TP\\
LVHIS 063 & ESO 383-G087    & 0 & 0 & TN\\
LVHIS 065 & NGC 5408        & 1 & 1 & TP\\
LVHIS 066 & Circinus Galaxy & 1 & 1 & TP\\
LVHIS 067 & UKS 1424-460    & 1 & 1 & TP\\
LVHIS 068 & ESO 222-G010    & 1 & 1 & TP\\
LVHIS 070 & ESO 272-G025    & 0 & 0 & TN\\
LVHIS 071 & ESO 223-G009    & 1 & 1 & TP\\
LVHIS 072 & ESO 274-G001    & 1 & 1 & TP\\
LVHIS 075 & IC 4662         & 1 & 1 & TP\\
LVHIS 076 & ESO 461-G036    & 1 & 1 & TP\\
LVHIS 077 & IC 5052         & 1 & 1 & TP\\
LVHIS 078 & IC 5152         & 1 & 1 & TP\\
LVHIS 079 & UGCA 438        & 0 & 0 & TN\\
LVHIS 080 & UGCA 442        & 1 & 1 & TP\\
LVHIS 081 & ESO 149-G003    & 1 & 1 & TP\\
LVHIS 082 & NGC 7793        & 1 & 1 & TP\\ \hline
\end{tabular}
\caption{LVHIS galaxies chosen from the LVHIS database as suitable for testing. The targets have beam major axes ranging from $5.3^{\prime\prime}$ to $34.7^{\prime\prime}$  with a mean of $13.2^{\prime\prime}$ and have pixels/beam values ranging from $5.25$ to $34.74$ with a median value of $12.78$. The channel widths are 4 $\text{km\,s}^{-1}$ bar one target which has a channel width of 8 $\text{km\,s}^{-1}$.  Of all 61 galaxies in the test set, 10 ($16\%$) were identified by eye as most likely to be classified as $\kappa < 0.5$ and their associated model predictions are shown. Of these 10 galaxies $8$ were correctly identified as low $\kappa$ by the binary classifier with no false negative predictions.}
\label{table:LVHIS}
\end{table*}

\bsp	
\label{lastpage}
\end{document}